\documentclass[a4paper,11pt,oneside]{article}

\usepackage[english]{babel}
\usepackage[T1]{fontenc}
\usepackage[utf8]{inputenc}

\usepackage[pdftex,unicode]{hyperref}
\hypersetup{pdftitle=Zero-Inflated Autoregressive Conditional Duration Model for Discrete Trade Durations with Excessive Zeros}
\hypersetup{pdfauthor=Francisco Blasques and Vladimír Holý and Petra Tomanová}

\usepackage{color}
\definecolor{mycolor}{rgb}{0.078,0.440,0.405}
\hypersetup{colorlinks=true, linkcolor=mycolor, anchorcolor=mycolor, citecolor=mycolor, filecolor=mycolor, urlcolor=mycolor}

\usepackage[margin=60pt]{geometry}
\usepackage[authoryear]{natbib}

\usepackage{amsmath}
\usepackage{amssymb}
\usepackage{graphicx}
\usepackage{booktabs}
\usepackage{multirow}
\usepackage[group-minimum-digits=3]{siunitx}

\usepackage{amsthm}
\newtheorem{lemma}{Lemma}
\newtheorem{proposition}{Proposition}

\begin{document}

\begin{center}
{\Large \bfseries Zero-Inflated Autoregressive Conditional Duration Model \\ for Discrete Trade Durations with Excessive Zeros}
\end{center}

\begin{center}
{\bfseries Francisco Blasques} \\
VU University Amsterdam and Tinbergen Institute \\
De Boelelaan 1105, NL-1081HV Amsterdam, The Netherlands \\
\href{f.blasques@vu.nl}{f.blasques@vu.nl}
\end{center}

\begin{center}
{\bfseries Vladimír Holý} \\
Prague University of Economics and Business \\
Winston Churchill Square 1938/4, 130 67 Prague 3, Czechia \\
\href{mailto:vladimir.holy@vse.cz}{vladimir.holy@vse.cz} \\
\end{center}

\begin{center}
{\bfseries Petra Tomanová} \\
Prague University of Economics and Business \\
Winston Churchill Square 1938/4, 130 67 Prague 3, Czechia \\
\href{mailto:petra.tomanova@vse.cz}{petra.tomanova@vse.cz}
\end{center}

\noindent
\textbf{Abstract:}
In finance, durations between successive transactions are usually modeled by the autoregressive conditional duration model based on a continuous distribution omitting zero values. Zero or close-to-zero durations can be caused by either split transactions or independent transactions. We propose a discrete model allowing for excessive zero values based on the zero-inflated negative binomial distribution with score dynamics. This model allows to distinguish between the processes generating split and standard transactions. We use the existing theory on score models to establish the invertibility of the score filter and verify that sufficient conditions hold for the consistency and asymptotic normality of the maximum likelihood of the model parameters. In an empirical study, we find that split transactions cause between 92 and 98 percent of zero and close-to-zero values. Furthermore, the loss of decimal places in the proposed approach is less severe than the incorrect treatment of zero values in continuous models.
\\

\noindent
\textbf{Keywords:}
Financial High-Frequency Data, Autoregressive Conditional Duration Model, Zero-Inflated Negative Binomial Distribution, Generalized Autoregressive Score Model.
\\

\noindent
\textbf{JEL Codes:}
C22, C41, C58.
\\

\section{Introduction}
\label{sec:intro}

%%% ACD models %%%
An important aspect of financial high-frequency data analysis is modeling of durations between various events. These include times of recording of transactions (trade durations), times when price changes by a given level (price durations), and times when volume reaches a given level (volume durations). Financial durations exhibit strong serial correlation, i.e.\ long durations are usually followed by long durations and short durations are followed by short durations. To capture this time dependence, \cite{Engle1998} proposed the autoregressive conditional duration (ACD) model.

%%% split transactions %%%
We focus on trade durations and one of their particular empirical characteristics -- the frequent occurrence of zero durations, i.e. trades executed at the same time. Zero durations are typically assumed to be caused by so-called split transactions, i.e.\ large trades broken into two or more smaller trades (see e.g.\ \citealp{Pacurar2008}). Subsequently, observations with the same timestamp are merged and the resulting prices are calculated as the average of prices weighted by volume. From the perspective of time series of trade durations, zero values are simply discarded. There is an obvious issue with this approach -- unrelated transactions that just occur at the same time but do not originate from the same source might be merged as well and their zero durations discarded. Nevertheless, this is the most common approach in the ACD literature dating back to \cite{Engle1998}. Dealing with zero values is even a necessity for ACD models based on distributions that do not contain zero in their support.
% as is the case of all distributions listed in Table \ref{tab:distributions} except the exponential distribution.
Alternatively, \cite{Bauwens2006} suggested setting zero durations to a small given value instead of discarding them. This transformation allows to keep all observations in the dataset but is quite arbitrary and distorts the distribution of durations near zero. From an economic point of view, however, it makes sense to consider split transactions as one single trade (see e.g.\ \citealp{Grammig2002}).

\begin{figure}
\begin{center}
\includegraphics[height=10cm]{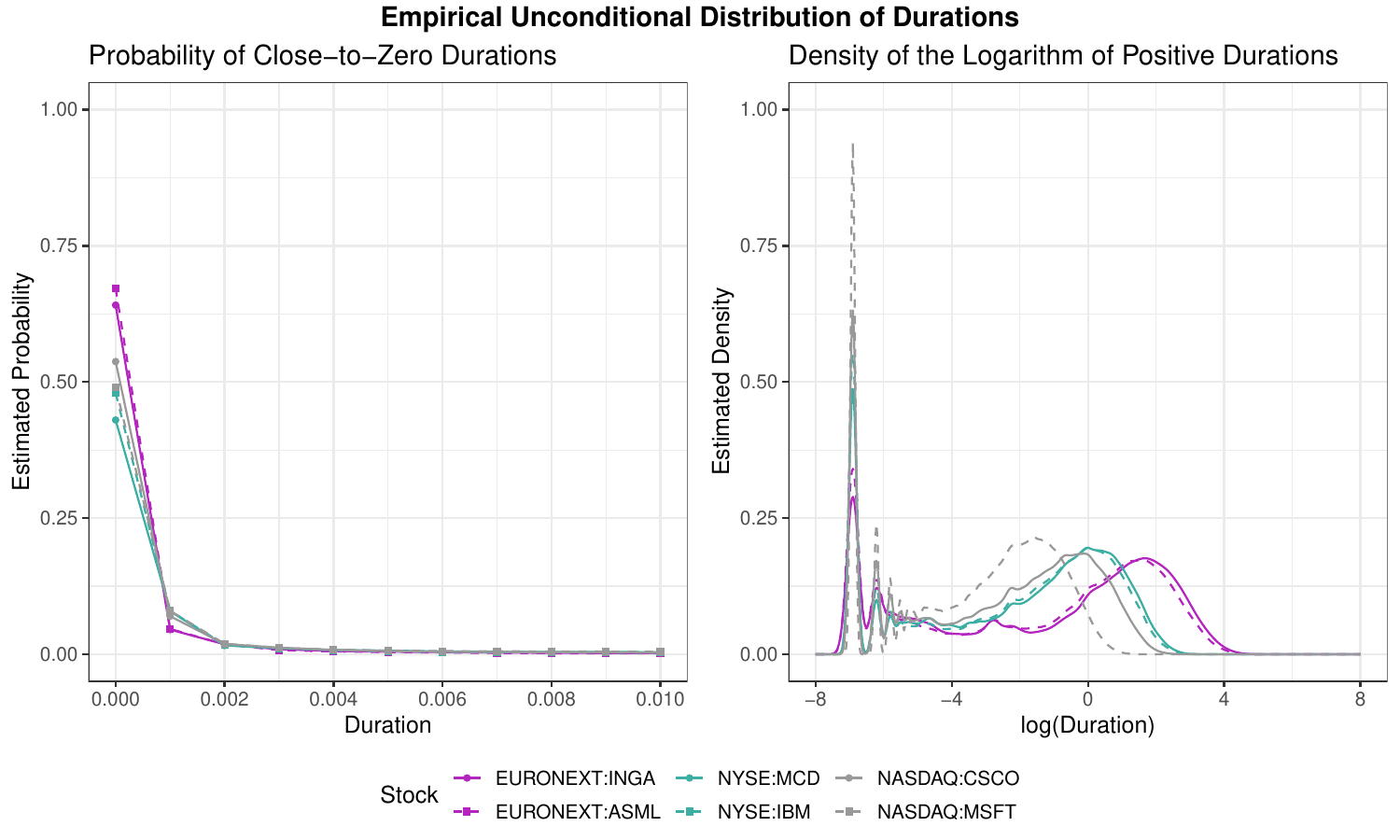}
\caption{The probability of durations between 0 and 0.01 seconds (left plot) and the density function of logarithmic durations estimated using the Gaussian kernel (right plot) in June, 2021. Zero durations are excluded.}
\label{fig:motiv}
\end{center}
\end{figure}

%%% timestamp precision %%%
Datasets analyzed by \cite{Engle1998} and others at the turn of the millennium had timestamps with precision to one second. Nowadays it is standard that transactions are recorded with precision to one millisecond, one microsecond, or even one nanosecond by some exchanges. This high detail causes an additional problem -- split transactions do not have to occur at the exact same time. An anecdotal evidence is presented in Table \ref{tab:motivSplit}. This has already been recognized e.g.\ by \cite{Grammig2002} who treated successive trades with either non-increasing or non-decreasing prices within one second as one large trade. Let us take a closer look at a recent dataset consisting of 6 stocks traded on the EURONEXT, NYSE, and NASDAQ exchanges with precision to one millisecond obtained from the Thomson Reuters database. The right plot of Figure \ref{fig:motiv} shows the density of the logarithm of durations estimated by the Parzen--Rosenblatt window method. Values equal to exactly zero are omitted from this figure. The density of log-durations is concentrated in two areas for each stock -- a ``hill'' in the middle of the plot and a ``wave'' in the left part of the plot. The ``wave'' shape is caused by discreteness of the data and captures durations close to zero. The left-most spike corresponds to 0.001 seconds, the next to it to the right to 0.002, and so on. For better readability of these close-to-zero durations, the left plot of Figure \ref{fig:motiv} shows their occurrence in data. First of all, we can see that exactly zero durations make up between 43 and 67 percent of all durations for the individual stocks. Durations equal to 0.001 are also quite frequent and make up between 5 and 8 percent. Durations equal to 0.002 make up about 2 percent and 0.003 durations about 1 percent. Other descriptive statistics are reported in Table \ref{tab:descr}. The main message here is that Figure \ref{fig:motiv} suggests that durations are generated by two processes -- one process generates dispersed values corresponding to unrelated transactions and the other process generates zero or close-to-zero values corresponding to split transactions.

%%% drawbacks of the traditional approach %%%
The traditional approach which assumes that all split transactions have exactly zero duration and all zero durations correspond to a split transaction is therefore not very suitable. Firstly, as mentioned above, discarding all zero durations might also discard zero durations corresponding to unrelated transactions. Secondly, and more importantly, keeping all positive durations might also keep close-to-zero durations corresponding to split transactions. Discarding all zeros and no positive values then leads to distorted distribution caused by inaccurate representation of values near zero.

%%% outline of our approach %%%
We propose to model durations by a mixture of two processes generating unrelated and split transactions respectively. We artificially reduce the precision of durations by rounding down the values to hundredths of a second, i.e.\ centiseconds, and operate within a discrete framework. With this reduced precision, we assume that all close-to-zero durations corresponding to split transactions fall into the new group of exactly zero durations, i.e.\ their original values are lower than 0.01 seconds. We then employ a zero-inflated distribution of \cite{Lambert1992} for modeling of durations. This distribution assumes that one process generates integer values greater or equal to zero and another process generates only zero values. The probability of unrelated transactions with zero durations is then determined by the distribution of positive values while the probability of split transactions with zero durations is given by the inflation parameter of the zero-inflated distribution. We are therefore able to estimate the ratio between unrelated and split transactions. In the empirical study, we demonstrate that the loss of precision of durations is redeemed by the simplicity of our model and its ability to accommodate for both unrelated and split transactions.

%%% proposed ZIACD model %%%
Given the discussion above, we propose in this paper a new zero-inflated autoregressive conditional duration (ZIACD) model. We base our model on the negative binomial distribution to accommodate for overdispersion in durations (see \citealp{Boswell1970, Cameron1986, Christou2014}). The excessive zero durations caused by split transactions are captured by the zero-inflated modification of the negative binomial distribution (see \citealp{Greene1994}). We let the scale, dispersion, and inflation parameters of the distribution be time-varying and follow the dynamics of generalized autoregressive score (GAS) models, also known as dynamic conditional score models (see \citealp{Creal2013, Harvey2013}). In the GAS framework, time-varying parameters are dependent on their lagged values and a scaled score of the conditional observation density. 

%%% contributions %%%
In this paper, we establish the invertibility of the GAS filter for the ZIACD model and the consistency and asymptotic normality of the maximum likelihood estimator for the case of time-varying scale parameter and static dispersion and zero-inflation parameters. In an empirical study of the stock market, we demonstrate that the proposed ZIACD model for durations rounded to centiseconds is usable in practice and is superior to continuous models with the incorrect treatment of zero values.

%%% paper structure %%%
The rest of the paper is structured as follows. In Section \ref{sec:lit}, we review the related literature on ACD and GAS models. In Section \ref{sec:model}, we propose the ZIACD model based on the zero-inflated negative binomial distribution. In Section \ref{sec:theory}, we verify the asymptotic properties of the maximum likelihood estimator for the case of time-varying scale. In Section \ref{sec:emp}, we describe characteristics of financial durations data, fit the proposed ZIACD model within a discrete framework, and compare it to a continuous model. In Section \ref{sec:disc}, we discuss the use of the proposed ZIACD model for low-precision data and alternative mixture ACD models as topics for future research. We conclude the paper in Section \ref{sec:conclusion}.

\begin{table}
\begin{center}
\caption{An excerpt from the limit order book of the MSFT stock on June 21, 2012.}
\label{tab:motivSplit}
\footnotesize
\begin{tabular}{llllrr}
\toprule
Message Time & Order ID & Event & Direction & Size & Price \\
\midrule
09:30:01.146 & 16333185 & Submission & Buy  &  300 & \$30.99 \\
\qquad \vdots \\
09:30:01.370 & 16333185 & Execution  & Buy  &  100 & \$30.99 \\
09:30:01.377 & 16333185 & Execution  & Buy  &  200 & \$30.99 \\
\qquad \vdots \\
09:30:03.550 & 16576783 & Submission & Sell & 3000 & \$30.99 \\
\qquad \vdots \\
09:30:03.553 & 16576783 & Execution  & Sell &  400 & \$30.99 \\
09:30:03.555 & 16576783 & Execution  & Sell &  400 & \$30.99 \\
09:30:03.555 & 16576783 & Execution  & Sell &  300 & \$30.99 \\
09:30:03.627 & 16576783 & Deletion   & Sell & 1900 & \$30.99 \\
\bottomrule
\end{tabular}
\end{center}
\end{table}

\section{Literature Review}
\label{sec:lit}

In this section, we examine two fundamental cornerstones of our paper: the Autoregressive Conditional Duration (ACD) model and the Generalized Autoregressive Score (GAS) model. These established models serve as the foundation for our novel contribution, the zero-inflated autoregressive conditional duration (ZIACD) model. 

\subsection{Autoregressive Conditional Duration Models}
\label{sec:litAcd}

Since the seminal paper of \cite{Engle1998}, many extensions of the original ACD model have been proposed in the literature. \cite{Bauwens2000} introduced the \emph{logarithmic ACD model} utilizing the logarithmic transformation and exogenous variables. Logarithmic model with a slightly different dynamic was considered by \cite{Lunde1999a}. Other proposed models include the \emph{fractionally integrated ACD model} of \cite{Jasiak1998}, \emph{threshold ACD model} of \cite{Zhang2001}, \emph{Box-Cox ACD model} of \cite{Hautsch2001,Hautsch2003}, \emph{asymmetric ACD model} of \cite{Bauwens2003}, \emph{additive and multiplicative ACD model} of \cite{Hautsch2011}, and \emph{directional ACD model} of \cite{Jeyasreedharan2014}. Time-varying and non-stationary ACD models were studied by \cite{Bortoluzzo2010} and \cite{Mishra2017}. Joint models for durations and prices were proposed by \cite{Engle2000}, \cite{Grammig2002}, \cite{Russell2005} and \cite{Herrera2013}. 

\cite{Ghysels2004a} proposed the \emph{stochastic volatility duration model}, which accounts for mean and variance dynamics in financial duration processes. Additionally, \cite{Bauwens2004a} introduced the \emph{stochastic conditional duration (SCD) model}, which was further extended by \cite{Feng2004} and \cite{Xu2011}. \cite{Feng2004} proposed the SCD model with leverage effect and \cite{Xu2011} added an interaction element between the duration process and the latent autoregressive process. \cite{Hujer2005} proposed \emph{Markov switching ACD model} that extends the traditional ACD model by introducing an unobservable stochastic process modeled by a Markov chain. \cite{Chen2013} proposed \emph{Markov-switching multifractal duration model}, which allows for modeling long memory in the duration process. \cite{Fernandes2006} developed a family of ACD models that encompasses most common specifications, where the nesting relies on a Box-Cox transformation.

Numerous studies in the literature also explore the incorporation of information about zero durations. \cite{Zhang2001} included an indicator of multiple transactions as an explanatory variable in their ACD model. \cite{Veredas2002} noticed that many simultaneous transactions occur at round prices suggesting many traders post limit orders to be executed at round prices -- this is an empirical phenomenon known as price clustering (see e.g. the literature review in \citealp{Holy2022b}). More recently, \cite{Liu2018} examined the effect of zero durations on integrated volatility estimation.

The first ACD models analyzed by \cite{Engle1998} utilize the exponential and Weibull distributions. However, since then, various continuous distributions have been employed in duration modeling; an overview can be found in Table \ref{tab:distributions}. Additionally, several studies in the literature have proposed ACD models based on mixtures of distributions. \cite{DeLuca2003} and \cite{DeLuca2004} suggested using a mixture of two exponential distributions to capture distinct behaviors of informed and uninformed traders. This work was further extended by \cite{DeLuca2009}, proposing the incorporation of the two exponential distributions with time-varying weights. On the other hand, to account for the unobserved market heterogeneity of traders, \cite{Gomez-Deniz2016a,Gomez-Deniz2017} proposed finite and infinite mixture of distributions based on non-exponentials, specifically a mixture of an inverse Gaussian distribution. For a survey of duration analysis, see \cite{Pacurar2008}, \cite{Bauwens2009}, \cite{Hautsch2011}, and \cite{Saranjeet2019}.

\begin{table}
\begin{center}
\caption{The use of continuous distributions in ACD models.}
\label{tab:distributions}
\footnotesize
\begin{tabular}{llr}
\toprule
Article            & Distribution           & Parameters \\
\midrule
\cite{Engle1998}   & Exponential            & 1 \\
\cite{Engle1998}   & Weibull                & 2 \\
\cite{Lunde1999a}  & Generalized Gamma      & 3 \\
\cite{Grammig2000} & Burr                   & 3 \\
\cite{Hautsch2001} & Generalized F          & 4 \\
\cite{Bhatti2010}  & Birnbaum--Saunders     & 2 \\
\cite{Xu2013}      & Log-Normal             & 2 \\
\cite{Leiva2014}   & Power-Exponential B--S & 3 \\
\cite{Leiva2014}   & Student's t B--S       & 3 \\
\cite{Zheng2016}   & Fréchet                & 2 \\
\bottomrule
\end{tabular}
\end{center}
\end{table}

\subsection{Generalized Autoregressive Score Models}
\label{sec:litGas}

\emph{Generalized autoregressive score (GAS)} models \citep{Creal2013}, also known as \emph{dynamic conditional score} models \citep{Harvey2013}, capture dynamics of time-varying parameters by the autoregressive term and the scaled score of the conditional observation density (see Section \ref{sec:modelGas} for further details). GAS models belong to the class of observation-driven models, as defined by \cite{Cox1981}, and thus have their advantages, e.g. observation-driven models can be estimated in a straightforward manner by the maximum likelihood method and their parameters are perfectly predictable given the past information. Moreover, \cite{Blasques2015} investigated information-theoretic optimality properties of the score function of the predictive likelihood and showed that only parameter updates based on the score will always reduce the local Kullback--Leibler divergence between the true conditional density and the model-implied conditional density. \cite{Koopman2016} find that observation-driven models based on the score perform comparably to parameter-driven models in terms of predictive accuracy.

The GAS specification includes many commonly used econometric models. For example, the GAS model with the normal distribution, the inverse of the Fisher information scaling and time-varying variance results in the GARCH model while the GAS model with the exponential distribution, the inverse of the Fisher information scaling and time-varying expected value results in the ACD model \citep{Creal2013}. The GAS framework can be utilized for discrete models as well. \cite{Koopman2018} used discrete copulas based on the Skellam distribution for high-frequency stock price changes. \cite{Koopman2019} used the bivariate Poisson distribution for a number of goals in football matches and the Skellam distribution for a score difference. \cite{Gorgi2018} used the Poisson distribution as well as the negative binomial distribution for offensive conduct reports. \cite{Holy2022b} used a mixture of double Poisson distributions to model price clustering in high-frequency prices.

\cite{Andres2012} specified ACD-like models belonging to the GAS framework and applied them to intra-day stock market data, considering both range and duration. \cite{Tomanova2021} utilized the GAS model based on the generalized gamma distribution in the spirit of ACD models, and demonstrated that this approach outperforms the traditional method that assumes times between arrivals follow the exponential distribution with a constant rate, making it a superior choice for modeling arrivals in queueing systems.

A comprehensive list of papers on GAS models can be found at \url{http://gasmodel.com}.

\section{Zero-Inflated ACD Model}
\label{sec:model}

%%% definition of duration process %%%
Let $T_0 \leq T_1 \leq \cdots \leq T_n$ be random variables denoting times of transactions. Trade durations are then defined as $X_i = T_i - T_{i-1}$ for $i = 1, \ldots, n$. As we operate in a discrete framework, we assume $T_i \in \mathbb{N}_0$, $i = 0, \ldots, n$ and $X_i \in \mathbb{N}_0$, $i = 1, \ldots, n$.\footnote{Note that this assumption is not restrictive since durations are naturally discrete and non-negative. Thus when expressed in the units corresponding to precision of the timestamps (e.g. seconds, milliseconds, \dots), the durations are natural numbers (with zero).} We further assume trade durations $X_i$ to follow some given discrete distribution with conditional probability mass function $P[X_i = x_i | \theta]$, where $x_i$ are observations and $\theta =(\theta_1, \ldots, \theta_l)'$ are parameters. First, we consider trade durations to follow the negative binomial distribution. Next, we extend the negative binomial distribution to capture excessive zeros using the zero-inflated model. Finally, we let parameters be time-varying with the generalized autoregressive score dynamics. 
%and denote them as $f_{i} = (f_{i,1} \ldots, f_{i,k})'$.  assume trade durations $X_i$ to follow a distribution with the time-varying parameters $f_{i}$ given by $P[X_i = x_i | f_{i}, \theta]$. 
%The model utilizes the \emph{score} for time-varying parameters $f_i$ defined as
%\begin{equation}
%\nabla(x_i, f_i) = \frac{\partial \ln \textrm{P}[X_i = x_i | f_i, \theta]}{\partial f_i}
%\end{equation}
%and the \emph{Fisher information} for time-varying parameters $f_i$ defined as
%\begin{equation}
%\mathcal{I}(f_i) = \mathrm{E} \Big[ \nabla(x_i, f_i) \nabla(x_i, f_i)' \Big| f_i, \theta \Big] = - \mathrm{E} \Bigg[ \frac{\partial^2 \ln \textrm{P}[X_i = x_i | f_i, \theta]}{\partial f_i \partial f_i'} \Bigg| f_i, \theta \Bigg].
%\end{equation}
%Note, that the latter equality requires some regularity conditions \citep{Lehmann1998}.

\subsection{Negative Binomial Distribution}
\label{sec:modelNegbin}

%%% use of negative binomial distribution %%%
Non-negative integer variables are commonly analyzed using count data models based on specific underlying distribution, most notably the Poisson distribution and the negative binomial distribution (see \citealp{Cameron2013}). A distinctive feature of the Poisson distribution is that its expected value is equal to its variance. This characteristic is too strict in many applications as count data often exhibit overdispersion, a higher variance than the expected value. A generalization of the Poisson distribution overcoming this limitation is the negative binomial distribution with one parameter determining its expected value and another parameter determining its excess dispersion.

%%% definition of negative binomial distribution %%%
The \emph{negative binomial (NB) distribution} can be derived in many ways (see \citealp{Boswell1970}). We use the NB2 parameterization of \cite{Cameron1986} derived from the Poisson-gamma mixture distribution. It is the most common parametrization used in the negative binomial regression according to \cite{Cameron2013}. The probability mass function with scale parameter $\mu > 0$ and dispersion parameter $\alpha \geq 0$ is
\begin{equation}
\label{eq:negbinProb}
\mathrm{P} [X_i = x_i | \mu, \alpha] = \frac{\Gamma (x_i + \alpha^{-1})}{\Gamma (x_i + 1) \Gamma (\alpha^{-1})} \left( \frac{\alpha^{-1}}{\alpha^{-1} + \mu} \right)^{\alpha^{-1}} \left( \frac{\mu}{\alpha^{-1} + \mu} \right)^{x_i}  \quad \text{for } x_i = 0,1,2,\ldots.
\end{equation}
The expected value and variance is
\begin{equation}
\label{eq:negbinMoments}
\begin{aligned}
\mathrm{E}[X_i] &= \mu, \\
\mathrm{var}[X_i] &= \mu (1 + \alpha \mu). \\
\end{aligned}
\end{equation}
%The score for the parameter $\mu$ is
%\begin{equation}
%\label{eq:negbinScore}
%\nabla (x_i, \mu) = \mu^{-1} (x_i - \mu)(\alpha \mu + 1)^{-1} \quad \text{for } x_i = 0,1,2,\ldots.
%\end{equation}
%The Fisher information for the parameter $\mu$ is
%\begin{equation}
%\label{eq:negbinFisher}
%\mathcal{I} (\mu) = \mu^{-1}(\alpha \mu + 1)^{-1}.
%\end{equation}
Special cases of the negative binomial distribution include the Poisson distribution for $\alpha = 0$ and the geometric distribution for $\alpha = 1$.

\subsection{Zero-Inflation}
\label{sec:modelZero}

%%% use of zero-inflated negative binomial distribution %%%
The zero-inflated distribution is an extension of a discrete distribution allowing the probability of zero values to be higher than the probability given by the original distribution. In the zero-inflated distribution, values are generated by two components -- one component generates only zero values while the other component generates integer values (including zero values) according to the original distribution. \cite{Lambert1992} proposed the zero-inflated Poisson model and \cite{Greene1994} used zero-inflated model for the negative binomial distribution.

%%% definition of zero-inflated negative binomial distribution %%%
The \textit{zero-inflated negative binomial distribution} is a discrete distribution with three parameters: scale parameter $\mu > 0$, dispersion parameter $\alpha \geq 0$ and probability of excessive zero values $\pi \in [0, 1)$. The variable $X_i$ follows the zero-inflated negative binomial distribution if
\begin{equation}
\begin{array}{ll}
X_i \sim 0 & \text{ with probability } \pi, \\
X_i \sim \mathrm{NB}(\mu, \alpha) & \text{ with probability } 1-\pi. \\
\end{array}
\end{equation}
The first process generates only zeros and corresponds to split transactions, while the second process generates values from the negative binomial distribution and corresponds to regular transactions. The probability mass function is
\begin{equation}
\label{eq:zinegbinProb}
\begin{aligned}
\mathrm{P} [X_i = 0 | \mu, \alpha, \pi] &= \pi + (1 - \pi)  \left( \frac{\alpha^{-1}}{\alpha^{-1} + \mu} \right)^{\alpha^{-1}}, \\
\mathrm{P} [X_i = x_i | \mu, \alpha, \pi] &= (1 - \pi) \frac{\Gamma (x_i + \alpha^{-1})}{\Gamma (x_i + 1) \Gamma (\alpha^{-1})} \left( \frac{\alpha^{-1}}{\alpha^{-1} + \mu} \right)^{\alpha^{-1}} \left( \frac{\mu}{\alpha^{-1} + \mu} \right)^{x_i} \quad \text{for } x_i = 1,2,\ldots. \\
\end{aligned}
\end{equation}
The expected value and variance is
\begin{equation}
\label{eq:zinegbinMoments}
\begin{aligned}
\mathrm{E}[X_i] &= \mu (1 - \pi), \\
\mathrm{var}[X_i] &= \mu(1 - \pi) (1 + \pi \mu + \alpha \mu). \\
\end{aligned}
\end{equation}
The score vector is given by
\begin{equation}
\nabla (x_i, \mu, \alpha, \pi) = 
\begin{pmatrix}
(\pi - 1) (\alpha \mu + 1)^{-1} \Big( 1 + \pi (\alpha \mu + 1)^{\alpha^{-1}} - \pi \Big)^{-1} \\
\alpha^{-2} \Big(\ln(\alpha \mu + 1) - \alpha \mu (\alpha \mu + 1)^{-1} \Big) \Big( 1 -  \pi (\pi - 1)^{-1} (\alpha \mu + 1)^{\alpha^{-1}} \Big)^{-1} \\
\Big( (\alpha \mu + 1)^{\alpha^{-1}} - 1 \Big) \Big( 1 + \pi (\alpha \mu + 1)^{\alpha^{-1}} - \pi \Big)^{-1} \\
\end{pmatrix}
\end{equation}
for $x_i = 0$ and
\begin{equation}
\nabla (x_i, \mu, \alpha, \pi) = 
\begin{pmatrix}
\mu^{-1} (x_i - \mu)(\alpha \mu + 1)^{-1} \\
\alpha^{-2} \Big(\ln(\alpha \mu + 1) + \alpha (x_i - \mu) (\alpha \mu + 1)^{-1} + \psi_0 (\alpha^{-1}) - \psi_0 (x_i + \alpha^{-1}) \Big) \\
(\pi - 1)^{-1} \\
\end{pmatrix}
\end{equation}
for $x_i = 1,2,\ldots$. 

%The score for the parameter $\mu$ is
%\begin{equation}
%\label{eq:zinegbinScore}
%\begin{aligned}
%\nabla (0, \mu) &= (\pi - 1) (\alpha \mu + 1)^{-1} \Big( 1 + \pi (\alpha \mu + 1)^{\alpha^{-1}} - \pi \Big)^{-1}, \\
%\nabla (x_i, \mu) &= \mu^{-1} (x_i - \mu)(\alpha \mu + 1)^{-1} \quad \text{for } x_i = 1,2,\ldots. \\
%\end{aligned}
%\end{equation}
%The Fisher information for the parameter $\mu_i$ is
%\begin{equation}
%\label{eq:zinegbinFisher}
%\mathcal{I} (\mu_i) = \frac{\pi(\pi - 1)}{(\alpha \mu_i + 1)^2 \Big( \pi (\alpha \mu_i + 1)^{\alpha^{-1}} - \pi + 1 \Big)} + \frac{1 -\pi}{\mu_i(\alpha \mu_i + 1)}.
%\end{equation}

%Special cases of the zero-inflated negative binomial distribution include the the negative binomial distribution for $\pi = 0$, zero-inflated Poisson distribution for $\alpha = 0$ and the zero-inflated geometric distribution for $\alpha = 1$.

\subsection{Score-Driven Dynamics}
\label{sec:modelGas}

%%% formulation of gas models %%%
\emph{Generalized autoregressive score (GAS)} models \citep{Creal2013} capture dynamics of time-varying parameters $\tilde{f}_{i} = (\tilde{f}_{i,1},\ldots,\tilde{f}_{i,k})'$ by the autoregressive term and the scaled score of the conditional observation density (or the conditional observation probability mass function in the case of discrete distribution). The time-varying parameters $\tilde{f}_i$ follow the recursion
\begin{equation}
\label{eq:gasRecursion}
\tilde{f}_{i+1} = C + B \tilde{f}_i + A S(\tilde{f}_i) \nabla(x_i, \tilde{f}_i),
\end{equation}
where $C = (c_1,\ldots,c_k)'$ are the constant parameters, $B = \mathrm{diag}(b_1,\ldots,b_k)$ are the autoregressive parameters, $A = \mathrm{diag}(a_1,\ldots,a_k)$ are the score parameters, $S(\tilde{f}_i)$ is the scaling function for the score and $\nabla(x_i, \tilde{f}_i)$ is the score. In the original paper of \cite{Creal2013}, authors noted that via the choice of the scaling function $S(\tilde{f}_i)$, the GAS model allows for additional flexibility in how the score is used for updating $\tilde{f}_i$. The commonly used scaling functions in the GAS literature are based on the Fisher information matrix. We explored this option, however, we have not found it very suitable for the GAS model with the negative binomial distribution since the Fisher information for the parameter $\alpha$ does not have a closed-form. Consequently, the approximation of the Fisher information brings undue computational complexity resulting in an overly time-consuming optimization procedure. In order to keep our model simple, from now on we avoid the scaling, which is also a widely used option in the GAS literature. Moreover, \cite{Holy2020} showed that the differences of models performance based on different scaling functions are negligible in the case of the negative binomial distribution.
%As the scaling function, we consider
%\begin{itemize}
%\item unit scaling, i.e.\ $S(f_i) = I$,
%\item square root of inverse of the Fisher information scaling, i.e.\ $S(f_i) = \mathcal{I}(f_i)^{-\frac{1}{2}}$,
%\item inverse of the Fisher information scaling, i.e.\ $S(f_i) = \mathcal{I}(f_i)^{-1}$.
%\end{itemize}
%Note that each scaling function results in a different GAS model. 

The long-term mean and unconditional value of the time-varying parameters is $\tilde{f}=(I-B)^{-1} C$. The parameters $\tilde{f}_i$ in \eqref{eq:gasRecursion} are assumed to be unbounded. However, some distributions require bounded parameters (e.g.\ variance greater than zero). The standard solution in the GAS framework is to use an unbounded parametrization $f_i = H(\tilde{f}_i)$, which follows the GAS recursion instead of the original parametrization $\tilde{f}_i$, i.e.
\begin{equation}\label{eq:gasProposedRecursion}
f_{i+1} = c + b f_i + a s(x_i, f_i),
\end{equation}
where $c$ are the constant parameters, $b$ are the autoregressive parameters, $a$ are the score parameters, and $s(x_i, f_i)$ is the reparametrized score. The reparametrized score equals to 
\begin{equation}
\label{eq:gasReparamScore}
s(x_i, f_i) = \dot{H}^{-1}(\tilde{f}_i) \nabla(x_i, \tilde{f}_i),
\end{equation}
where $\dot{H}(\tilde{f}_i) = \partial H(\tilde{f}_i) / \partial \tilde{f}_i'$ is the derivation of $H(\tilde{f}_i)$. 
%while the Fisher information of the reparametrized model equals to
%\begin{equation}
%\label{eq:gasReparamFisher}
%\tilde{\mathcal{I}}(\tilde{f}_i) = \dot{H}'^{-1}(f_i) \mathcal{I}(f_i) \dot{H}^{-1}(f_i),
%\end{equation}
%where $\dot{H}(f_i) = \partial H(f_i) / \partial f_i'$ is the derivation of $H(f_i)$.

\subsection{Zero-Inflated Autoregressive Conditional Duration Model}
\label{sec:modelProposed}

%%% proposed gas model %%%
We consider a model where observations follow the zero-inflated negative binomial distribution with the time-varying scale parameter $\mu_i$, time-varying dispersion parameter $\alpha_i$ and time-varying inflation parameter $\pi_i$ specified in \eqref{eq:zinegbinProb}. We use an unbounded parametrization with the exponential link for the scale and dispersion parameters and logistic transformation for the inflation parameter, i.e. $f_i = (\ln (\mu_{i}),\ln (\alpha_{i}),\ln (\pi_{i}/(1-\pi_{i})))' $. Parameters $f_i$ are assumed to follow the recursion in \eqref{eq:gasProposedRecursion}, where the score for the zero-inflated negative binomial distribution is given by
\begin{equation}
\label{eq:zinegbinScoreZero}
s(x_i, f_i) = 
\begin{pmatrix}
\mu_i(\pi_i - 1) (\alpha_i \mu_i + 1)^{-1} \Big( 1 + \pi_i (\alpha_i \mu_i + 1)^{\alpha_i^{-1}} - \pi_i \Big)^{-1} \\
\alpha_i^{-1} \Big(\ln(\alpha_i \mu_i + 1) - \alpha_i \mu_i (\alpha_i \mu_i + 1)^{-1} \Big) \Big( 1 -  \pi_i (\pi_i - 1)^{-1} (\alpha_i \mu_i + 1)^{\alpha_i^{-1}} \Big)^{-1} \\
\pi_i(1-\pi_i) \Big( (\alpha_i \mu_i + 1)^{\alpha_i^{-1}} - 1 \Big) \Big( 1 + \pi_i (\alpha_i \mu_i + 1)^{\alpha_i^{-1}} - \pi_i \Big)^{-1} \\
\end{pmatrix}
\end{equation}
for $x_i = 0$ and
\begin{equation}
\label{eq:zinegbinScorePositive}
s(x_i, f_i) = 
\begin{pmatrix}
(x_i - \mu_i)(\alpha_i \mu_i + 1)^{-1} \\
\alpha_i^{-1} \Big(\ln(\alpha_i \mu_i + 1) + \alpha_i (x_i - \mu_i) (\alpha_i \mu_i + 1)^{-1} + \psi_0 (\alpha_i^{-1}) - \psi_0 (x_i + \alpha_i^{-1}) \Big) \\
-\pi_i \\
\end{pmatrix}
\end{equation}
for $x_i = 1,2,\ldots$.

\section{Estimation and Asymptotic Properties}
\label{sec:theory}

In this section, we focus on the model with the time-varying scale parameter $\mu_i$ and static dispersion $\alpha$ and inflation $\pi$ parameters. As such we set $f_i = \ln (\mu_{i})$ and $\theta=(\alpha,\pi,c,b,a)'$. The score in \eqref{eq:zinegbinScoreZero} and \eqref{eq:zinegbinScorePositive} simplifies to
\begin{equation}
\label{eq:gasProposedScore}
\begin{aligned}
s(0,f_i) &= \frac{(\pi - 1) \exp(f_{i}) }{(\alpha \exp(f_{i}) + 1) \left( 1 + \pi (\alpha \exp(f_{i}) + 1)^{\alpha^{-1}} - \pi \right)}, \\
s(x_{i}, f_i) &= \frac{x_i - \exp(f_{i})}{\alpha \exp(f_{i}) + 1} \quad \text{for }  x_i = 1,2,\ldots.
\end{aligned}
\end{equation}
For this GAS model with dynamics defined in \eqref{eq:gasProposedRecursion} and \eqref{eq:gasProposedScore}, we establish the invertibility of the score filter and verify that sufficient conditions hold for the consistency and asymptotic normality of the maximum likelihood of the model parameters.

The static parameter vector $\theta$ is estimated by the method of maximum likelihood
\begin{equation}
\label{eq:durationsEstLikelihood}
\hat{\theta}_{n} \in \arg\max_{\theta \in \Theta} \hat{L}_{n}(\theta),
\end{equation}
where $\hat{L}_{n}(\theta)$ denotes the log likelihood function obtained from a sequence of $n$ observations $x_1,\ldots,x_n$, which depends on the filtered time-varying parameter $\hat{f}_{1}(\theta),...,\hat{f}_{n}(\theta)$. Since we are dealing with observation-driven filters which require an initialization value $\hat{f}_{1}$, we make an important distinction here between $\hat{L}_{n}(\theta)$ and $L_{n}(\theta)$. The first log likelihood is a function of the filtered parameter $\hat{f}_{1}(\theta),...,\hat{f}_{n}(\theta)$ initialized at a given value $\hat{f}_{1}$. The second likelihood is a function of the filtered parameter $f_{1}(\theta),...,f_{n}(\theta)$ initialized at the true unobserved value $f_{1}$. Of course, since $f_{1}$ is unobserved, we typically have that $\hat{f}_{1} \neq f_{1}$. In practice, the sample log likelihood is thus given by 
\begin{equation}
\hat{L}_{n}(\theta) = \frac{1}{n} \sum_{i=1}^{n} \hat{\ell}_{i}(x_{i},\theta) = \frac{1}{n} \sum_{i=1}^{n} \ln \textrm{P}[X_i = x_i | \hat{f}_{i}(\theta),\theta].  
\end{equation}
In our case, the log likelihood is based on the zero-inflated negative binomial distribution
\begin{equation}
\begin{aligned}
\label{eq:loglik}	
\ln \textrm{P}[X_i = 0 | \hat{f}_{i}(\theta),\theta] &= \ln \left( \pi + (1 - \pi)  \left( \frac{\alpha^{-1}}{\alpha^{-1} + \mu_i} \right)^{\alpha^{-1}} \right), \\
\ln \textrm{P}[X_i = x_i | \hat{f}_{i}(\theta),\theta] &= \ln (1 - \pi) + \ln  \frac{\Gamma (x_i + \alpha^{-1})}{\Gamma (x_i + 1) \Gamma (\alpha^{-1})} + \frac{1}{\alpha} \ln \left( \frac{\alpha^{-1}}{\alpha^{-1} + \exp(\hat{f}_i)} \right) \\
& \quad  + x_{i} \ln  \left( \frac{\exp(\hat{f}_i)}{\alpha^{-1} + \exp(\hat{f}_i)} \right) \quad \text{for } x_i = 1,2,\ldots.
\end{aligned}
\end{equation}
Below, we show that the maximum likelihood estimator of the ZIACD model is consistent and asymptotically normal. The proof follows the structure laid down in \cite{Blasques2022}, but we focus on the particular case of discrete data $\{x_{i}\}_{i \in \mathbb{N}}$ with a probability mass function $\textrm{P}[X_i = x_i | f_{i}(\theta),\theta]$. In contrast, \cite{Blasques2022} treat a general case for continuous data with a smooth probability density function.

\subsection{Filter Invertibility}
\label{sec:theoryInvert}

Filter invertibility is crucial for statistical inference in the context of observation-driven time-varying parameter models; see e.g.\ \cite{Straumann2006}, \cite{Wintenberger2013}, and \cite{Blasques2022}. The filter $\{\hat{f}_{i}(\theta)\}_{i\in \mathbb{N}}$ initialized at some point $\hat{f}_{1} \in \mathbb{R}$ is said to be invertible if $\hat{f}_{i}(\theta)$ converges almost surely exponentially fast to a unique limit strictly stationary and ergodic sequence $\{f_{i}(\theta)\}_{i\in \mathbb{Z}}$,
\begin{equation*}
|\hat{f}_{i}(\theta)-f_{i}(\theta)| \stackrel{eas}{\to} 0 \quad \text{as} \quad  i \to \infty. 
\end{equation*} 
Let $L_{n}(\theta)$ denote the log likelihood which depends on the limit time-varying parameter $f_{1}(\theta),...,f_{n}(\theta)$
\begin{equation*}
L_{n}(\theta) = \frac{1}{n} \sum_{i=1}^{n} \ell_{i}(x_{i},\theta) = \frac{1}{n} \sum_{i=1}^{n} \ln \textrm{P}[X_i = x_i | f_{i}(\theta),\theta],
\end{equation*}
and let $L_{\infty}$ denote the limit log likelihood function
\begin{equation*}
L_{\infty}(\theta) = \mathrm{E} [ \ell_{i}(\theta) ] = \mathrm{E} \left[ \ln \textrm{P}[X_i = x_i | f_{i}(\theta),\theta] \right].
\end{equation*}

Proposition \ref{prop:invertibility} appeals to the results in \cite{Blasques2022} to establish the invertibility of the score filter with zero-inflated negative binomial distribution as stated in \eqref{eq:gasProposedRecursion} and \eqref{eq:gasProposedScore}. The proof presented in Technical Appendix A is an application of the results in \cite{Blasques2022} to our current model. 

\begin{proposition}[Filter invertibility]
\label{prop:invertibility}
Consider the score-driven model with zero-inflated negative binomial distribution in \eqref{eq:gasProposedRecursion} and \eqref{eq:gasProposedScore}.
Let the observed data $\{x_{i}\}_{i\in \mathbb{N}}$ be strictly stationary and ergodic, with a logarithmic moment $\mathrm{E} [\ln^{+}|x_{i}|] < \infty$, and  let $\Theta$ be a compact parameter space defined as 	$$\Theta=[\alpha^{-},\alpha^{+}]\cdot [\pi^{-},\pi^{+}]\cdot [c^{-},c^{+}]\cdot  [b^{-},b^{+}]\cdot [a^{-},a^{+}]$$ and satisfying the following restrictions
	\begin{equation*}
	\begin{aligned}
	\frac{a^{+}(\pi^{-}-1)^{2}}{2 \alpha^{-}} +  \frac{a^{+}|\pi^{-}-1|}{(\alpha^{-})^{2}} + b^{+} &< 1, \\
	\mathrm{E}_{x_{i}>0} \left[ \ln \left( \frac{a^{+} (\alpha^{+} x_{i}+1) }{4\alpha^{-}} + b^{+} \right) \right]  &< 0. \\
	\end{aligned}
	\end{equation*}
Then the filter $\{\hat{f}_{i}(\theta)\}_{i \in \mathbb{N}}$  defined as $\hat{f}_{i+1} = c + b \hat{f}_i + a s(x_i, \hat{f}_i)$ is invertible, uniformly in $\theta \in \Theta$. 
\end{proposition}

\subsection{Consistency}
\label{sec:theoryConst}

Proposition \ref{prop:invertibility} gives us sufficient elements to characterize the asymptotic behavior of the ML estimator. This section uses existing theory on score models in \cite{Blasques2022} to verify   the strong consistency of the ML estimator $\hat{\theta}_{n}$ as the sample size $n$ diverges to infinity. 

For completeness, Lemma \ref{theo:consistency} states conditions for the consistency of the ML estimator. A sketch of the proof is offered in Technical Appendix A, and appropriate references are offered. This theorem naturally uses the invertibility properties established in Proposition \ref{prop:invertibility} for our zero-inflated negative binomial score model. Following \cite{Blasques2022}, this theorem allows for potential model mispecification. 

\begin{lemma}[Consistency of the ML estimator]
\label{theo:consistency}
Let the conditions of Proposition \ref{prop:invertibility} hold. Suppose further that the observed data has one bounded moment $\mathrm{E}[x_{i}]<\infty$, and  let  $\theta_{0}$ be the unique maximizer of the limit log likelihood function $\mathrm{E}[\ell_{i}(x_{i},\cdot)]: \Theta \to \mathbb{R}$  over the parameter space $\Theta$. Then  $\hat{\theta}_{n} \stackrel{as}{\to} \theta_{0} \in \Theta$ as $n \to \infty$.
\end{lemma}

\subsection{Asymptotic Normality}
\label{sec:theoryNorm}

Finally, we shed some light on the $\sqrt{n}$-consistency rate of $\hat{\theta}_{n}$ and the asymptotic normality of the standardized estimator $\sqrt{n}(\hat{\theta}_{n}-\theta_{0})$ as $n \to \infty$,  when the model is well specified.
For completeness, Lemma \ref{theo:normality} summarizes standard conditions for asymptotic normality.  A sketch of the proof is presented in Technical Appendix A, and we refer to \citet{Blasques2022} for additional details.  

\begin{lemma}[Asymptotic normality of the ML estimator]
\label{theo:normality}
Let the conditions of Lemma \ref{theo:consistency} hold. Suppose that the observed data has four bounded moments $\mathrm{E}|x_{i}|^{4}<\infty$, and  let the true parameter lie in the interior of the parameter space, i.e.~$\theta_{0} \in \mathrm{int}(\Theta)$. Finally, let the further regularity conditions stated in Theorem 4.16 of \citet{Blasques2022} hold. 
Then the ML estimator is asymptotically Gaussian  
\begin{equation*}
\sqrt{n}(\hat{\theta}_{n} - \theta_{0}) \stackrel{d}{\to} N(0,\mathcal{I}(\theta_{0})^{-1})	 \quad \text{as} \quad n \to \infty,
\end{equation*}
where $\mathcal{I}(\theta_{0})^{-1}$ denotes the inverse Fisher information matrix.
\end{lemma}

\section{Empirical Study}
\label{sec:emp}

\subsection{Data Overview}
\label{sec:empData}

%%% analyzed data %%%
In our empirical study, we analyze transaction data extracted from the Thomson Reuters Eikon. Eikon provides access to real-time market data and also contains historical intraday transactions. The data are taken from June to July of 2021. We analyze 6 stocks: ING Groep (INGA) and ASML Holding (ASML) which are listed on EURONEXT; McDonald's Corporation (MCD) and International Business Machines Corporation (IBM) which are listed on NYSE; Cisco Systems, Inc. (CSCO) and Microsoft Corporation (MSFT) which are listed on NASDAQ. 

We clean data using the following procedure. First, we delete observations with the timestamp outside the main trading hours and trading days. Second, for EURONEXT stocks, we delete all observations with the timestamps equal to the first timestamp of the day that occurs between 09:00:00 and 09:00:30. The reason is that the opening uncrossing (resulting from the morning auction) randomly occurs between 09:00:00 and 09:00:30. Third, we round the timestamp to the right precision (i.e. milliseconds) to fix the incorrect representation of the float.\footnote{For all analyzed stocks we observed that the sorted unique duration values are: 0, 0.000999927520751953, 0.00100016593933105, 0.00199985504150391, 0.00200009346008301, \dots. The Thomson Reuters data are stamped with precision to one millisecond and this strange behavior is caused by an issue related to the representation of the float, which can be easily fixed by rounding.} 

The statistical characteristics for cleaned data are presented in Table~\ref{tab:descr}. The two analyzed stocks listed on the NASDAQ belong to the most liquid stocks, while the stocks listed on the EURONEXT represent the least liquid stocks in our dataset. In June 2021, exact zero durations range from 43.01 percent (MCD) to 67.19 percent (ASML) and durations lower than 1 second form up to 98.57 percent (MSFT) of the dataset. For further descriptive statistics, see Table \ref{tab:descr}. The ZIACD and continuous models in the empirical study are estimated by the \texttt{gasmodel} package in R.

\begin{table}
\begin{center}
\caption{Descriptive statistics of trade durations in June and July, 2021.}
\label{tab:descr}
\footnotesize
\begin{tabular}{llrrrrrr}
\toprule
& & \multicolumn{2}{c}{EURONEXT} & \multicolumn{2}{c}{NYSE} & \multicolumn{2}{c}{NASDAQ} \\
\cmidrule(l{3pt}r{3pt}){3-4} \cmidrule(l{3pt}r{3pt}){5-6} \cmidrule(l{3pt}r{3pt}){7-8}
Statistic & Sample & INGA & ASML & MCD & IBM & CSCO & MSFT \\ 
\midrule
\multirow{2}{*}{\% = 0}        & June &  64.11 &  67.19 &  43.01 &  47.97 &  53.73 &  49.05 \\ 
                               & July &  57.78 &  65.66 &  46.01 &  48.75 &  54.14 &  48.93 \\ \\
\multirow{2}{*}{\% $<$ 0.01}   & June &  73.70 &  76.30 &  56.98 &  61.63 &  66.86 &  63.86 \\ 
                               & July &  67.52 &  74.53 &  59.78 &  63.01 &  67.11 &  63.93 \\ \\
\multirow{2}{*}{\% $<$ 0.1}    & June &  77.53 &  79.77 &  65.02 &  68.81 &  74.82 &  77.81 \\ 
                               & July &  71.86 &  78.50 &  67.18 &  71.13 &  74.48 &  79.20 \\ \\
\multirow{2}{*}{\% $<$ 1}      & June &  82.31 &  84.73 &  82.91 &  85.72 &  91.37 &  98.57 \\ 
                               & July &  78.37 &  85.11 &  84.59 &  88.75 &  90.72 &  99.05 \\ \\
\multirow{2}{*}{Mean}          & June &   1.56 &   1.19 &   0.58 &   0.47 &   0.26 &   0.10 \\ 
                               & July &   1.72 &   0.91 &   0.52 &   0.37 &   0.29 &   0.08 \\ \\
\multirow{2}{*}{Variance}      & June &  27.85 &  18.69 &   1.90 &   1.43 &   0.54 &   0.05 \\ 
                               & July &  26.01 &  10.31 &   1.72 &   1.02 &   0.63 &   0.04 \\ \\
\multirow{2}{*}{Std. Dev.}     & June &   5.28 &   4.32 &   1.38 &   1.19 &   0.73 &   0.23 \\ 
                               & July &   5.10 &   3.21 &   1.31 &   1.01 &   0.79 &   0.20 \\ \\
\multirow{2}{*}{95\%-Quantile} & June &   9.94 &   7.50 &   3.25 &   2.70 &   1.60 &   0.54 \\ 
                               & July &  10.48 &   5.66 &   2.96 &   2.14 &   1.73 &   0.46 \\ \\
\multirow{2}{*}{Obs. per Min.} & June &  38.48 &  50.50 & 103.55 & 128.53 & 227.47 & 622.69 \\ 
                               & July &  34.88 &  66.00 & 115.11 & 163.22 & 210.39 & 723.17 \\ \\
\multirow{2}{*}{Total Obs.}    & June & \num{431441} & \num{566303} & \num{888400} & \num{1102742} & \num{1951673} & \num{5342645} \\ 
                               & July & \num{391156} & \num{740150} & \num{942707} & \num{1336712} & \num{1641075} & \num{5922788} \\ 
\bottomrule
\end{tabular}
\end{center}
\end{table}

\subsection{In-Sample Performance}
\label{sec:empIn}

We use the proposed ZIACD model based on the zero-inflated negative binomial distribution with the time-varying scale, dispersion, and zero inflation parameters to fit observed durations rounded down to hundredths of a second using data from June 2021. The estimated coefficients are reported in Table \ref{tab:fitZiacdCoef}. All coefficients are significant at any reasonable level and their standard deviations are virtually zero due to huge sample sizes ranging from 431463 (INGA) to 5342667 (MSFT). We, therefore, report only the estimated values. The numbers of observations per minute are also reported in Table \ref{tab:descr}. As expected, the coefficient controlling the impact of the score $a$ is positive for all three parameters and all six stocks. This means that the score serves as a correction term that adjusts the time-varying parameters for the observed values. The autoregressive coefficient $b$ is also positive and quite high for all three parameters and all six stocks. In the case of the scale parameter, it is very close to one signaling high persistence of the time series.

Table \ref{tab:fitZiacdAvg} reports the average values of the scale, dispersion, and zero inflation parameters over time. Note that the average scale parameter (adjusted to seconds) is much higher than the sample mean reported in Table \ref{tab:descr} as our model is able to separate zeros attributed to split transactions which subsequently do not affect the scale parameter. On average, between 53.27 percent (MCD) and 74.88 percent (ASML) of all durations are excessive zeros generated by split transactions depending on the stock. This corresponds to the ratio of excessive zeros to all zeros ranging between 91.81 percent (MSFT) and 98.13 percent (ASML). In other words, between 1.87 percent (ASML) and 8.19 percent (MSFT) of zero durations are generated by unrelated transactions which should not be discarded from the data.

Table \ref{tab:fitZiacdAvg} also evaluates the fit of the ZIACD model. The mean absolute error is between 0.11 seconds (MSFT) and 2.50 seconds (INGA) while the root mean square error is between 0.21 (MSFT) and 5.22 (INGA). These values are quite high when compared to the predicted value $\mu_i (1 - \pi_i)$, on which both errors are based, with its mean ranging from 0.09 seconds (MSFT) to 1.58 seconds (INGA). This is caused by the fact that the predicted value is not very representative of the whole distribution as, on average, between 53.27 percent (MCD) and 74.88 percent (ASML) of all values are exactly zero while the rest have expected value between 0.22 seconds (MSFT) and 5.68 seconds (INGA). It is therefore more suitable to assess the fit of the model based on the whole distribution.

We focus on the probability of zeros given by the model. Table \ref{tab:fitZiacdAvg} reports the mean probabilities of zero value given by the model when the observed value is indeed zero and when the observed value is positive. For the INGA and ASML stocks, the difference between these two probabilities is lower than one percent suggesting a limited benefit of the dynamics in the zero-inflation parameter. For the more traded stocks, the difference is between 5.78 percent (MCD) and 9.58 percent (CSCO) suggesting a certain degree of predictive ability of the zero-inflation dynamics.

\begin{figure}
\begin{center}
\includegraphics[height=10cm]{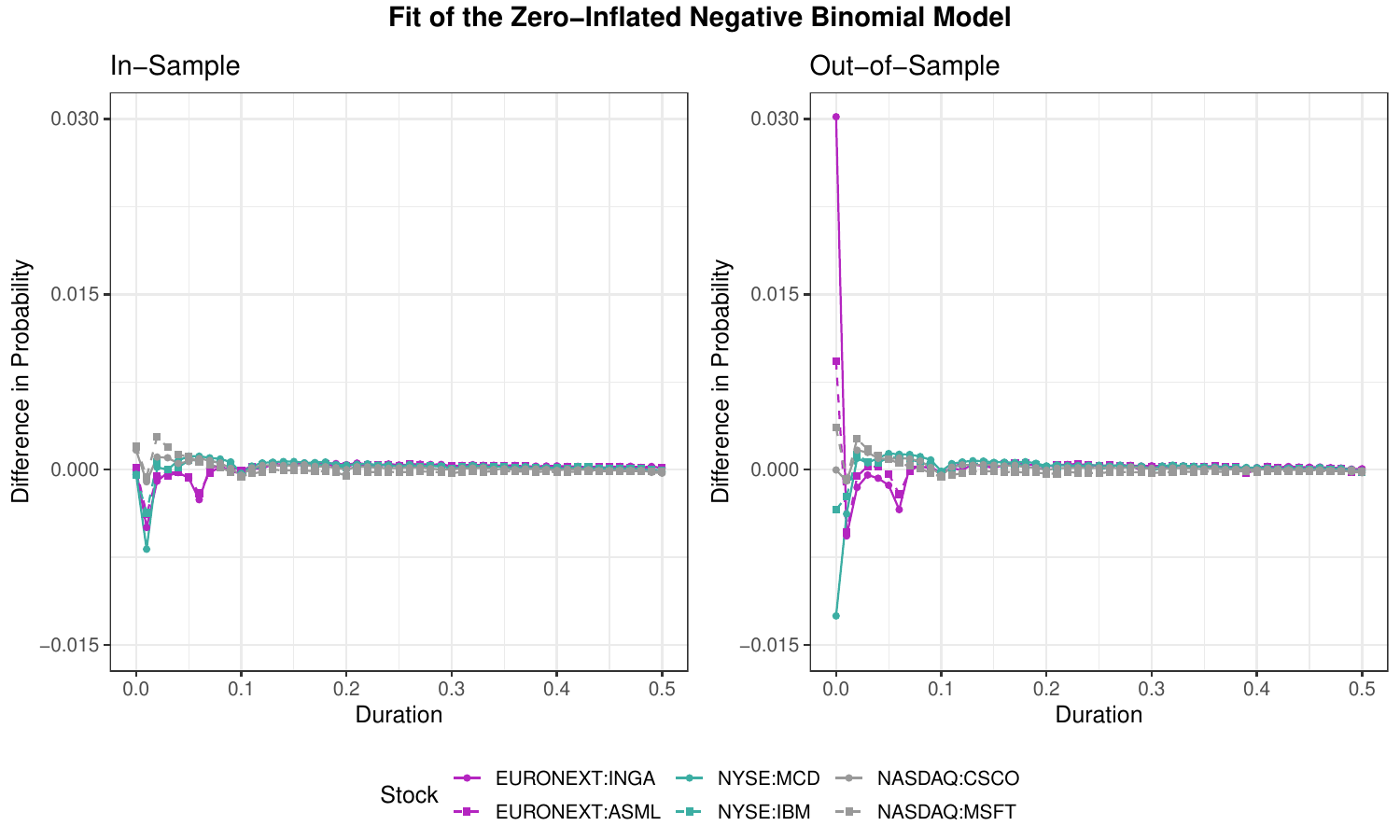}
\caption{The in-sample and out-of-sample difference between the conditional probabilities given by the zero-inflated negative binomial model and the unconditional distribution of observations.}
\label{fig:probZinegbin}
\end{center}
\end{figure}

The left plot of Figure \ref{fig:probZinegbin} studies the fit of the model in more detail by comparing the average conditional probabilities given by the ZIACD model with the unconditional empirical distribution. The largest deviation is -0.68 percent at 0.01 seconds for the MCD stock. This deviation is rather small but uncovers a systematic error as the probability of 0.01 durations is underestimated for all stocks. Similar underestimation is also present at 0.06 seconds for the ASML and INGA stocks traded on the EURONEXT exchange and at 0.10 seconds for all stocks. The latter two anomalies are also visible in the right plot of Figure \ref{fig:motiv} at -2.81 and -2.30 log-durations. The proposed model is therefore incorrectly specified and the true distribution of durations is much more complex. Nevertheless, the deviations of the conditional ZIACD probabilities are quite small and the model is usable in practice.

\begin{table}
\begin{center}
\caption{The estimated coefficients of the zero-inflated negative binomial model.}
\label{tab:fitZiacdCoef}
\footnotesize
\begin{tabular}{llrrrrrr}
\toprule
& & \multicolumn{2}{c}{EURONEXT} & \multicolumn{2}{c}{NYSE} & \multicolumn{2}{c}{NASDAQ} \\
\cmidrule(l{3pt}r{3pt}){3-4} \cmidrule(l{3pt}r{3pt}){5-6} \cmidrule(l{3pt}r{3pt}){7-8}
Parameter & Coef. & INGA & ASML & MCD & IBM & CSCO & MSFT \\ 
\midrule
               & $c$ & 0.006068 & 0.002591 & 0.000011 & 0.000151 & 0.000180 & 0.000064 \\
Scale          & $a$ & 0.109420 & 0.089377 & 0.032544 & 0.032434 & 0.051552 & 0.032155 \\
               & $b$ & 0.998958 & 0.999509 & 0.999996 & 0.999954 & 0.999913 & 0.999938 \\ \\
               & $c$ & 0.006364 & 0.061190 & 0.148700 & 0.129161 & 0.042488 & 0.000869 \\
Dispersion     & $a$ & 0.057713 & 0.216293 & 0.289589 & 0.243921 & 0.136988 & 0.021367 \\
               & $b$ & 0.992826 & 0.927438 & 0.806245 & 0.815294 & 0.948153 & 0.998387 \\ \\
               & $c$ & 0.030722 & 0.017910 & 0.048158 & 0.138758 & 0.116703 & 0.119207 \\
Zero Inflation & $a$ & 0.164058 & 0.100389 & 2.129143 & 2.177550 & 2.672883 & 2.542853 \\
               & $b$ & 0.968110 & 0.983785 & 0.680476 & 0.668047 & 0.856934 & 0.743213 \\
\bottomrule
\end{tabular}
\end{center}
\end{table}

\begin{table}
\begin{center}
\caption{The mean scale parameter (in seconds), the mean dispersion parameter, the mean inflation parameter (in percent), the mean ratio of zeros caused by split transactions (in percent), the mean predicted value (in seconds), the mean absolute error (in seconds), the root mean square absolute error (in seconds), the mean probabilities of zero value given by the zero-inflated negative binomial model when the observation is either zero or positive (in percent), and the mean log-likelihood.}
\label{tab:fitZiacdAvg}
\footnotesize
\begin{tabular}{lrrrrrr}
\toprule
 & \multicolumn{2}{c}{EURONEXT} & \multicolumn{2}{c}{NYSE} & \multicolumn{2}{c}{NASDAQ} \\
\cmidrule(l{3pt}r{3pt}){2-3} \cmidrule(l{3pt}r{3pt}){4-5} \cmidrule(l{3pt}r{3pt}){6-7}
Variable & INGA & ASML & MCD & IBM & CSCO & MSFT \\ 
\midrule
Mean Scale                          &  5.68 &  4.89 &  1.28 &  1.14 &  0.70 &  0.22 \\ 
Mean Dispersion                     &  2.45 &  2.35 &  2.17 &  2.02 &  2.26 &  1.70 \\ 
Mean Zero Inflation                 & 72.06 & 74.88 & 53.27 & 58.60 & 63.01 & 58.63 \\ 
Mean Split Ratio                    & 97.78 & 98.13 & 93.49 & 95.09 & 94.23 & 91.81 \\ 
Mean Predicted Value                &  1.58 &  1.22 &  0.60 &  0.48 &  0.27 &  0.09 \\
Mean Absolute Error                 &  2.50 &  1.96 &  0.72 &  0.59 &  0.31 &  0.11 \\ 
Root Mean Square Error              &  5.22 &  4.28 &  1.29 &  1.11 &  0.66 &  0.21 \\ 
$\mathrm{P} [X_i = 0]$ When $x_i=0$ & 67.48 & 67.97 & 65.60 & 66.47 & 69.15 & 69.66 \\ 
$\mathrm{P} [X_i = 0]$ When $x_i>0$ & 67.09 & 67.74 & 59.81 & 60.62 & 59.57 & 61.52 \\ 
Mean Log-Likelihood                 & -2.42 & -2.17 & -3.01 & -2.69 & -2.16 & -2.00 \\ 
\bottomrule
\end{tabular}
\end{center}
\end{table}

\subsection{Out-of-Sample Performance}
\label{sec:empOut}

In this section, we use the models estimated using durations from June 2021 and perform one-step-ahead forecasts during July 2021 to assess their long-term behavior. The right plot of Figure \ref{fig:probZinegbin} shows deviations of the average out-of-sample conditional probabilities given by the ZIACD model from the unconditional empirical distribution. Similarly to the left plot of Figure \ref{fig:probZinegbin}, the probabilities at 0.01, 0.06, and 0.10 seconds are systematically underestimated. However, the highest deviations are in the case of the probabilities of zero durations. The difference in probability reaches 3.02 percent (INGA) and drops down to -1.25 percent (MCD). This is related to a change in the occurrence of zero values in July. According to Table \ref{tab:descr}, the unconditional probability of zero values decreases from 64.11 to 57.78 percent for the INGA stock while it increases from 43.01 to 46.01 percent for the MCD stock. Note that the other descriptive statistics in Table \ref{tab:descr} also change considerably.

%This does not, however, translate to a significant decrease in the log-likelihood. As depicted in Figure \ref{fig:daily}, the log-likelihood exhibits strong day-to-day variations, both in-sample and out-of-sample; but there is no apparent trend, which is also supported by a simple linear regression. This means that although the model is not capable of capturing long-term changes in the process, it remains relatively robust to them. 

%Nevertheless, the proposed model is suitable only for short-term predictions. If the goal is to capture changing characteristics of durations, employing a non-stationary model could be more appropriate. We leave an analysis of long-term dynamics of excessive zero probability as a topic for future research.

\begin{figure}
\begin{center}
\includegraphics[height=10cm]{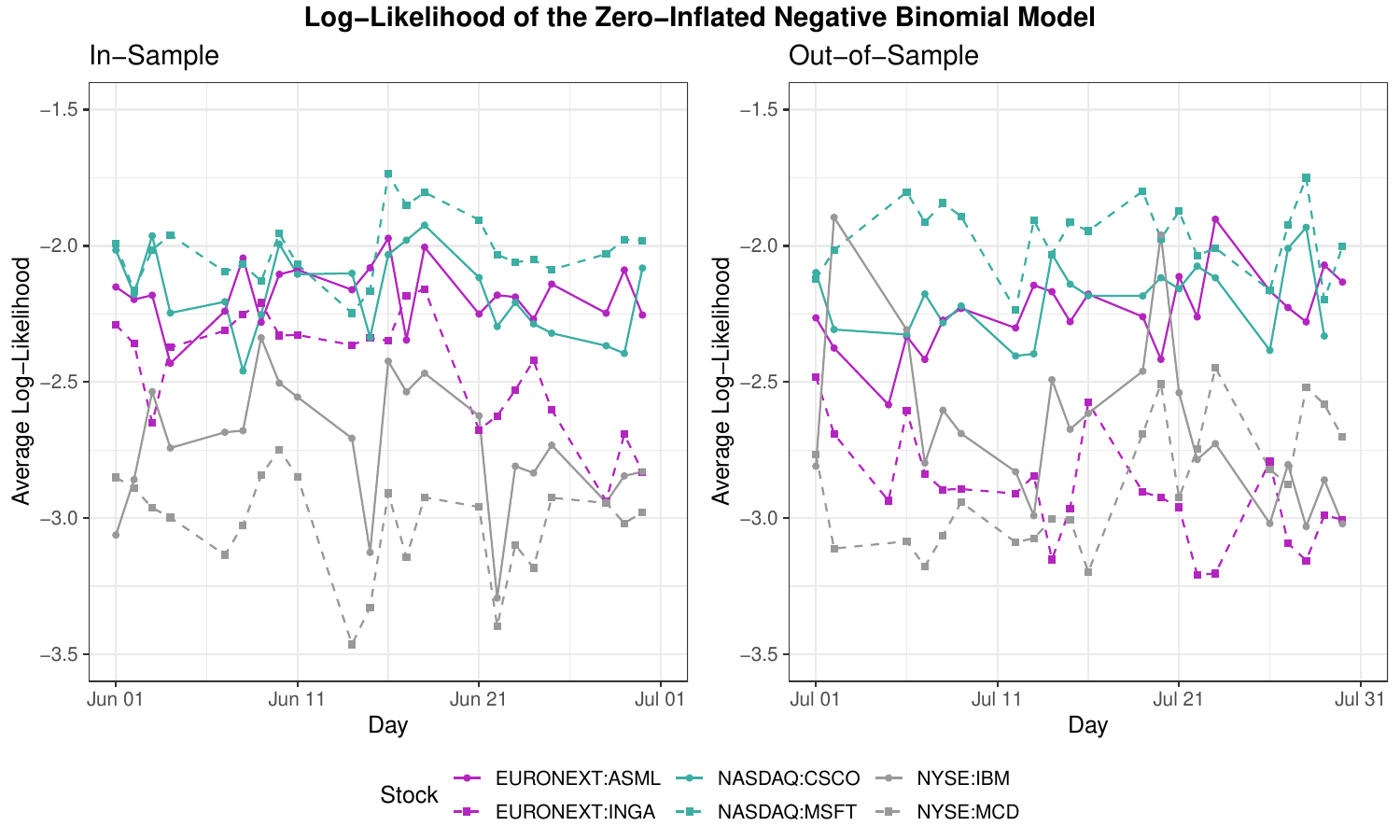}
\caption{The in-sample and out-of-sample average daily log-likelihood of the zero-inflated negative binomial model.}
\label{fig:daily}
\end{center}
\end{figure}

However, this does not translate to a significant decrease in the log-likelihood. Figure \ref{fig:daily} shows no apparent trend in the out-of-sample average daily log-likelihood, which is further supported by a simple linear regression analysis. This indicates that while the model may not be capable of accurately predicting long-term changes in the process, its forecasting performance does not significantly deteriorate over the long run. Furthermore, it should be noted that despite the overall stable performance, there is a noticeable volatility in day-to-day changes in the log-likelihood. This suggests that the accuracy of forecasts can vary significantly from one day to another. 

To summarize, the proposed model is best suited for short-term predictions. For capturing changing characteristics of durations, it would be more appropriate to use a non-stationary model. As for the long-term dynamics of excessive zero probability, we leave this analysis as a topic for future research.

\subsection{Model Specification}
\label{sec:empSpec}

We compare the proposed ZIACD model, which is based on the zero-inflated negative binomial distribution and has all three parameters time-varying, with models imposing some restrictions. Specifically, Table \ref{tab:fitCompDynamics} compares models based on the Poisson, geometric, and negative binomial distributions together with their zero-inflated versions. All parameters in these models are time-varying. On the other hand, Table \ref{tab:fitCompDistribution} compares models based on the zero-inflated negative binomial distribution with some parameters static and some time-varying. We use two criteria to compare the models -- the difference in the Akaike information criterion (AIC) for the in-sample fit and the Diebold-Mariano (DM) statistic for the out-of-sample fit. When comparing two models, a positive difference in the AIC favors the second model over the first model while a positive value of the DM statistic favors the first model over the second model. The DM statistic has asymptotically the standard normal distribution under the null hypothesis of equivalent out-of-sample log-likelihoods. More details on these criteria are given in Technical Appendix B. Not surprisingly in such large datasets, the most general specification of the model has the best fit. We do not report $p$-values for the DM statistic as it is significant at any reasonable level in all cases due to huge sample sizes.

There is clear evidence of overdispersion, i.e.\ the variance higher than the expected value. According to Table \ref{tab:fitZiacdAvg}, the average value of the dispersion parameter $\alpha$ in the zero-inflated negative binomial model ranges between $1.70$ (MSFT) and $2.45$ (INGA). This favors the negative binomial distribution over the Poisson distribution with fixed $\alpha = 0$ and the geometric distribution with fixed $\alpha = 1$. Overdispersion is also supported by the difference in the AIC and the DM statistic reported in Table \ref{tab:fitCompDistribution}. The Poisson distribution has the highest AIC for all stocks while the geometric distribution has the worst DM statistic compared to the zero-inflated negative binomial distribution. One possible reason for overdispersion could just be the presence of excessive zeros. Indeed, the zero-inflated Poisson and geometric distributions perform better than their original versions. However, they are still inferior to the zero-inflated negative binomial distribution suggesting that there is overdispersion present in non-zero values as well. Table \ref{tab:fitCompDynamics} further shows that the specification with the time-varying dispersion parameter performs significantly better than the static one. This improvement of the in-sample and out-of-sample fit is, however, the smallest among all specifications in Tables \ref{tab:fitCompDistribution} and \ref{tab:fitCompDynamics}. For some smaller data samples of less traded assets or with shorter periods of time (such as a day), the model with static dispersion parameter might be more suitable due to possible overfitting.

Our analysis also reveals the presence of excessive zeros suggesting the existence of the process generating only zero values (i.e.\ split transactions) alongside the process generating regular durations. According to Table \ref{tab:fitZiacdAvg}, the average probability of excessive zeros $\pi$ in the zero-inflated negative binomial model ranges between 53.27 percent (MCD) and 74.88 percent (ASML). The presence of excessive zeros is further supported by the better in-sample and out-of-sample fit for the zero-inflated distributions as reported in Table \ref{tab:fitCompDistribution}. Table \ref{tab:fitCompDynamics} shows that it is also suitable to let the zero-inflation parameter be time-varying as this increases the in-sample and out-of-sample fit, particularly for the more traded stocks MCD, IBM, CSCO, and MSFT. This is in line with the mean probabilities of zero value when the observation is either zero or positive reported in Table \ref{tab:fitZiacdAvg}.

\begin{table}
\begin{center}
\caption{The difference in the Akaike information criterion (AIC) and the Diebold--Mariano (DM) statistic for the models based on the Poisson distribution (P), the geometric distribution (G), the negative binomial distribution (NB), the zero-inflated Poisson distribution (ZIP), the zero-inflated geometric distribution (ZIG), and the zero-inflated negative binomial distribution (ZINB).}
\label{tab:fitCompDistribution}
\footnotesize
\begin{tabular}{llrrrrrr}
\toprule
& & \multicolumn{2}{c}{EURONEXT} & \multicolumn{2}{c}{NYSE} & \multicolumn{2}{c}{NASDAQ} \\
\cmidrule(l{3pt}r{3pt}){3-4} \cmidrule(l{3pt}r{3pt}){5-6} \cmidrule(l{3pt}r{3pt}){7-8}
Distribution & Crit. & INGA & ASML & MCD & IBM & CSCO & MSFT \\ 
\midrule
\multirow{2}{*}{P / ZINB}   & AIC & 267461915.62 & 280418352.77 & 138404692.48 & 148394239.86 & 142150126.37 & 124884854.32 \\
                            & DM  &      -235.37 &      -269.95 &      -384.02 &      -436.68 &      -361.32 &      -945.72 \\ \\
\multirow{2}{*}{G / ZINB}   & AIC &   2995721.03 &   3894926.03 &   3283604.41 &   4270769.74 &   6100642.18 &  10641774.31 \\
                            & DM  &      -689.62 &      -981.64 &      -793.38 &      -953.95 &      -668.61 &     -1272.63 \\ \\
\multirow{2}{*}{NB / ZINB}  & AIC &     45891.96 &     58214.58 &    118521.51 &    153740.09 &    279306.75 &    617624.25 \\
                            & DM  &      -119.12 &      -138.71 &      -221.39 &      -261.55 &      -287.93 &      -469.99 \\ \\
\multirow{2}{*}{ZIP / ZINB} & AIC & 104329981.44 & 100918549.37 &  60299913.57 &  58956197.68 &  58305592.35 &  43332973.41 \\
                            & DM  &      -170.40 &      -210.26 &      -291.16 &      -311.60 &      -336.45 &      -617.43 \\ \\
\multirow{2}{*}{ZIG / ZINB} & AIC &     49991.79 &     50910.51 &     84210.77 &     77231.38 &    122581.75 &    112468.26 \\
                            & DM  &       -80.79 &       -97.10 &      -108.19 &      -113.01 &      -123.58 &      -144.40 \\                   
\bottomrule
\end{tabular}
\end{center}
\end{table}

\begin{table}
\begin{center}
\caption{The difference in the Akaike information criterion (AIC) and the Diebold--Mariano (DM) statistic for the zero-inflated negative binomial model with all parameters static (SSS), dynamic $\mu$ (DSS), dynamic $\mu$, $\alpha$ (DDS), dynamic $\mu$, $\pi$ (DSD), and dynamic $\mu$, $\alpha$, $\pi$ (DDD).}
\label{tab:fitCompDynamics}
\footnotesize
\begin{tabular}{llrrrrrr}
\toprule
& & \multicolumn{2}{c}{EURONEXT} & \multicolumn{2}{c}{NYSE} & \multicolumn{2}{c}{NASDAQ} \\
\cmidrule(l{3pt}r{3pt}){3-4} \cmidrule(l{3pt}r{3pt}){5-6} \cmidrule(l{3pt}r{3pt}){7-8}
Dynamics & Crit. & INGA & ASML & MCD & IBM & CSCO & MSFT \\ 
\midrule
\multirow{2}{*}{SSS / DDD} & AIC & 21492.36 & 27667.68 & 264677.50 & 325382.37 & 839689.51 & 1780799.16 \\
                           & DM  &   -86.74 &  -133.22 &   -302.72 &   -393.75 &   -425.24 &    -762.36 \\ \\
\multirow{2}{*}{DSS / DDD} & AIC &  8135.80 &  7612.61 & 219595.15 & 271386.43 & 742989.86 & 1447522.61 \\
                           & DM  &   -57.16 &   -71.11 &   -274.29 &   -321.08 &   -396.20 &    -619.05 \\ \\
\multirow{2}{*}{DDS / DDD} & AIC &  5652.62 &  4831.46 &  86375.82 & 112311.74 & 212495.87 &  521863.12 \\
                           & DM  &   -50.55 &   -56.63 &   -177.80 &   -208.29 &   -231.05 &    -420.17 \\ \\
\multirow{2}{*}{DSD / DDD} & AIC &  1617.11 &  1953.18 &   7286.14 &   5069.21 &   6614.27 &    9203.83 \\
                           & DM  &   -13.99 &   -24.88 &    -32.19 &    -35.19 &    -24.04 &     -32.57 \\
\bottomrule
\end{tabular}
\end{center}
\end{table}

\subsection{Degree of Rounding}
\label{sec:empRound}

The choice of rounding to hundredths of a second, i.e.\ centiseconds, is motivated by Figure \ref{fig:motiv} which shows that the majority of excessive close-to-zero durations is concentrated in values 0 and 0.001 and the occurrence of larger values quickly decreases. In this section, we study the impact of different degrees of rounding and whether this choice is appropriate. Again, we use the difference in the AIC to assess the in-sample fit and the DM statistic to assess the out-of-sample fit. When comparing two models with different degrees of rounding, we compute the log-likelihood (which is the base for both AIC and DM) with respect to the rounding to fewer decimal places. A probability under the rounding to fewer decimal places is then the sum of the corresponding probabilities under the rounding to more decimal places. We consider rounding to zero decimal places (seconds), one decimal place (deciseconds), two decimal places (centiseconds), and three decimal places (milliseconds), i.e.\ the original data.

Table \ref{tab:fitCompRounding} shows the impact of increasing rounding. The rounding to centiseconds is clearly preferred over no rounding, i.e.\ precision to milliseconds. This is caused by the inability of the ZIACD model on milliseconds to account for an excessive probability of durations between 0.001 and 0.009 seconds; mostly, however, 0.001 seconds. The choice between the rounding to centiseconds and deciseconds differs for the individual stocks. For the INGA and AMSL stocks traded on the EURONEXT exchange, the model on deciseconds performs better. The difference in the AIC and the value of the DM statistic suggesting deciseconds are significant but smaller compared to the other values in Table \ref{tab:fitCompRounding}. To keep the reported results simple, we stick with the model on centiseconds. For the more trade stocks MCD, IBM, CSCO, and MSFT, the model on centiseconds clearly outperforms the model on deciseconds. Finally, deciseconds are preferred over seconds for all stocks.

\begin{table}
\begin{center}
\caption{The difference in the Akaike information criterion (AIC) and the Diebold--Mariano (DM) statistic for the zero-inflated negative binomial model based on data rounded to milliseconds (ms), centiseconds (cs), deciseconds (ds), and seconds (s).}
\label{tab:fitCompRounding}
\footnotesize
\begin{tabular}{llrrrrrr}
\toprule
& & \multicolumn{2}{c}{EURONEXT} & \multicolumn{2}{c}{NYSE} & \multicolumn{2}{c}{NASDAQ} \\
\cmidrule(l{3pt}r{3pt}){3-4} \cmidrule(l{3pt}r{3pt}){5-6} \cmidrule(l{3pt}r{3pt}){7-8}
Precision & Crit. & INGA & ASML & MCD & IBM & CSCO & MSFT \\ 
\midrule
\multirow{2}{*}{ms / cs} & AIC & 28004.18 & 37387.09 &  49082.83 &  58914.43 &  42215.78 &  178862.15 \\
                         & DM  &   -74.72 &  -102.18 &   -101.74 &    -97.79 &    -32.09 &    -114.94 \\ \\
\multirow{2}{*}{cs / ds} & AIC &  3085.80 &  4326.46 & -18588.85 & -28803.35 & -66676.58 & -233072.75 \\
                         & DM  &   -11.59 &   -29.46 &     47.73 &     76.70 &     87.26 &     218.46 \\ \\
\multirow{2}{*}{ds / s}  & AIC & -1071.15 & -1021.23 & -34331.20 & -42690.00 & -70647.10 &  -34868.11 \\
                         & DM  &    10.75 &    12.33 &     87.23 &     97.62 &    105.13 &      66.52 \\       
\bottomrule
\end{tabular}
\end{center}
\end{table}

\subsection{Comparison to Continuous Models}
\label{sec:empCont}

We compare the proposed discrete ZIACD model with continuous models based on the generalized gamma distribution (see Technical Appendix C) with GAS dynamics. The generalized gamma distribution contains the exponential, Weibull, and gamma distributions as special cases and belongs to the family of the generalized F distribution. The use of the generalized gamma distribution in ACD models was proposed by \cite{Lunde1999a}. Both \cite{Bauwens2004} and \cite{Fernandes2005} found that the generalized gamma distribution is more adequate than the exponential, Weibull, and Burr distributions. The study \cite{Xu2013} shows that the log-normal distribution does not outperform the generalized gamma distribution either. For these reasons, the generalized gamma distribution is our main candidate for the competing continuous distribution. In our comparison, we do not consider the generalized F distribution as it has four parameters and in most cases of financial durations reduces to the generalized gamma distribution as discussed by \cite{Hautsch2003} and \cite{Hautsch2011}. We also do not consider the Birnbaum--Saunders distribution as it models the median instead of the scale parameter and therefore does not strictly belong to the traditional ACD class. Models based on continuous distributions must address the issue of zero durations. We consider two ways of dealing with zero values in continuous models -- discarding them and truncating them to a given value. Furthermore, we consider three values for truncating -- 0.001, 0.0005, and 0.0001 seconds. \cite{Bauwens2006} used truncation to the half of the smallest increment, which is 0.0005 seconds in our case. Similarly to the previous section, we compute log-likelihood on a discrete grid of centiseconds. In the case of discarding zeros, we compare the generalized gamma model with the zero-inflated negative binomial model that is also estimated without zero values.

\begin{figure}
\begin{center}
\includegraphics[height=10cm]{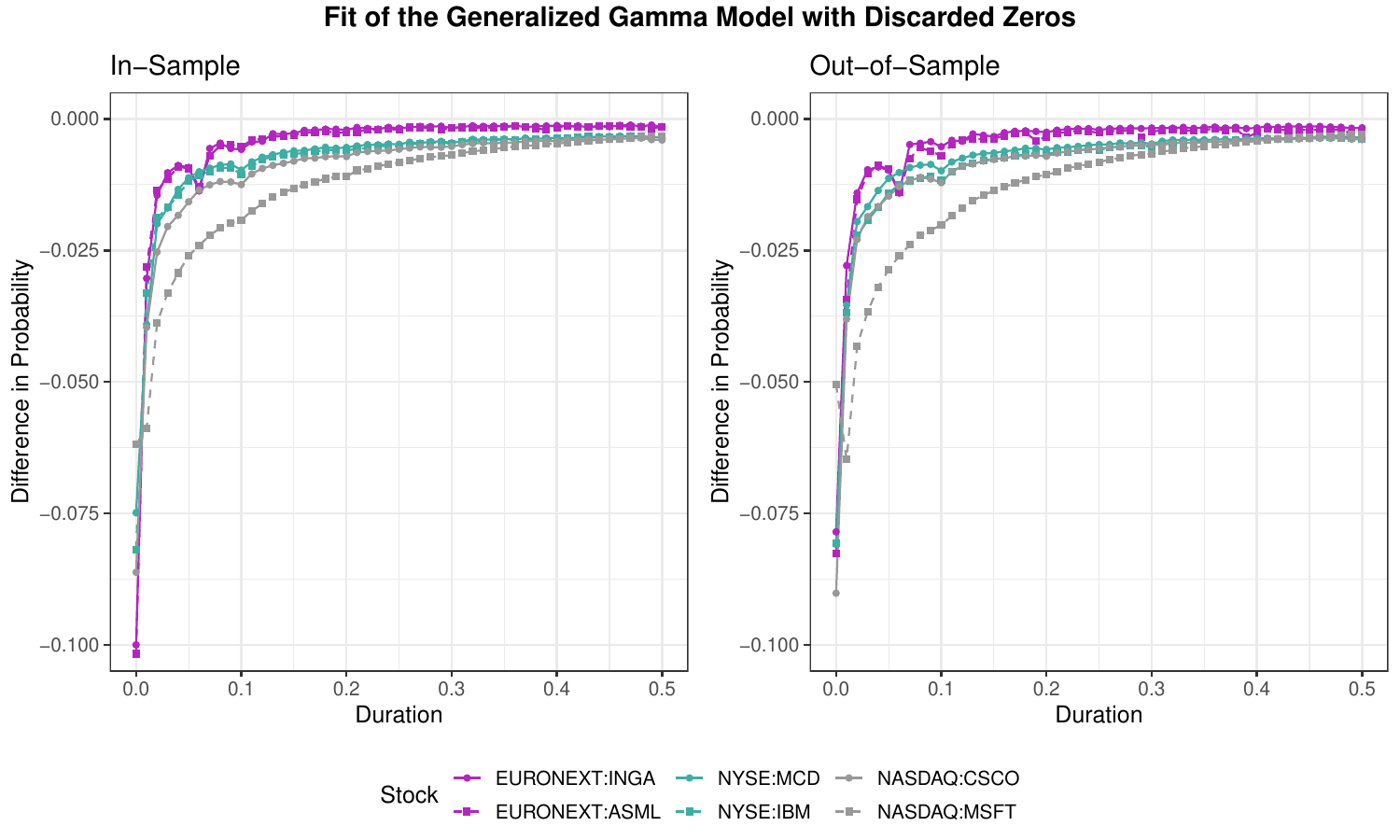}
\caption{The in-sample and out-of-sample difference between the conditional probabilities given by the generalized gamma model with discarded zeros and the unconditional distribution of observations.}
\label{fig:probDiscard}
\end{center}
\end{figure}

Figure \ref{fig:probDiscard} demonstrates the unsuitability of the approach discarding zeros. Similarly to Figure \ref{fig:probZinegbin}, the generalized gamma model is not able to capture unusually increased occurrence of 0.06 seconds (for the INGA and ASML stocks) and 0.10 seconds (for all stocks). A crucial problem, however, is significantly underestimated probabilities in the wider vicinity of zero. In the case of the zero value itself, the difference in probability reaches -10.17 percent for the AMSL stock. Note that Figure \ref{fig:probDiscard} has much larger scale than Figure \ref{fig:probZinegbin}. Table \ref{tab:fitCompGengamma} then confirms the superiority of the ZIACD model over the continuous alternatives in terms of the difference in the AIC and the DM statistic. Concerning the treatment of zero values, we can see that it is better to truncate zeros to smaller values but it is even better to just discard them. Either way, the results imply that the loss of decimal places in the proposed ZIACD model is of much less importance than the incorrect treatment of zero values in the continuous models.

\begin{table}
\begin{center}
\caption{The difference in the Akaike information criterion (AIC) and the Diebold--Mariano (DM) statistic for the generalized gamma model (GG) with zeros discarded (Discard) or truncated (Trunc) and the zero-inflated negative binomial model (ZINB).}
\label{tab:fitCompGengamma}
\footnotesize
\begin{tabular}{llrrrrrr}
\toprule
& & \multicolumn{2}{c}{EURONEXT} & \multicolumn{2}{c}{NYSE} & \multicolumn{2}{c}{NASDAQ} \\
\cmidrule(l{3pt}r{3pt}){3-4} \cmidrule(l{3pt}r{3pt}){5-6} \cmidrule(l{3pt}r{3pt}){7-8}
Model & Crit. & INGA & ASML & MCD & IBM & CSCO & MSFT \\ 
\midrule
\multirow{2}{*}{GG Discard / ZINB Discard} & AIC &  14808.54 &  18250.42 &  24298.98 &  32947.27 &  69568.44 &  143388.69 \\
                                           & DM  &    -41.41 &    -48.97 &    -72.66 &    -81.80 &   -103.98 &    -143.54 \\ \\
\multirow{2}{*}{GG Trunc to 0.001 / ZINB}  & AIC & 268465.60 & 356891.51 & 360261.19 & 469852.61 & 730365.26 & 1358026.51 \\
                                           & DM  &   -293.92 &   -385.96 &   -342.79 &   -387.75 &   -258.53 &    -548.24 \\ \\
\multirow{2}{*}{GG Trunc to 0.0005 / ZINB} & AIC & 218050.18 & 302587.56 & 325396.64 & 406946.28 & 572429.68 & 1111215.05 \\
                                           & DM  &   -245.36 &   -337.90 &   -203.65 &   -342.76 &   -365.74 &    -479.62 \\ \\
\multirow{2}{*}{GG Trunc to 0.0001 / ZINB} & AIC & 174616.63 & 227173.46 & 283440.84 & 329196.87 & 462917.51 & 1073765.49 \\
                                           & DM  &   -224.66 &   -271.87 &   -154.87 &   -299.34 &   -191.52 &    -461.16 \\                                      
\bottomrule
\end{tabular}
\end{center}
\end{table}

\section{Discussion}
\label{sec:disc}

\subsection{Discreteness of Data}
\label{sec:discRound}

As mentioned above, our paper studies data with high-precision timestamps. Although it is nowadays quite common that exchanges record transactions with precision to one millisecond or higher, one can encounter preprocessed datasets with precision to one second due to their easier readability. In some cases, this can even be the only dataset provided by the exchange to the public\footnote{For example, the Prague Stock Exchange currently records times of transactions with precision to one millisecond and distributes millisecond data to its members and external agencies. However, data provided to individuals have a precision of one second only.}. For these low-precision data, it is more natural to use a discrete model such as ours rather than a continuous model.

To our knowledge, \cite{Grimshaw2005} is the only paper addressing the issue of rounding in financial durations analysis. They found that ignoring the discreteness of data leads to a distortion of time-dependence tests in financial durations. More loosely related, \cite{Schneeweiss2010} reviewed the bias-inducing effects of rounding. \cite{Tricker1984a} and \cite{Taraldsen2011} explored the effects of rounding on the exponential distribution while \cite{Tricker1992} dealt with the gamma distribution. \cite{Zhang2010} and \cite{Li2011} found that the rounding errors in autoregressive processes can further accumulate making continuous models unreliable.

\begin{figure}
\begin{center}
\includegraphics[height=8cm]{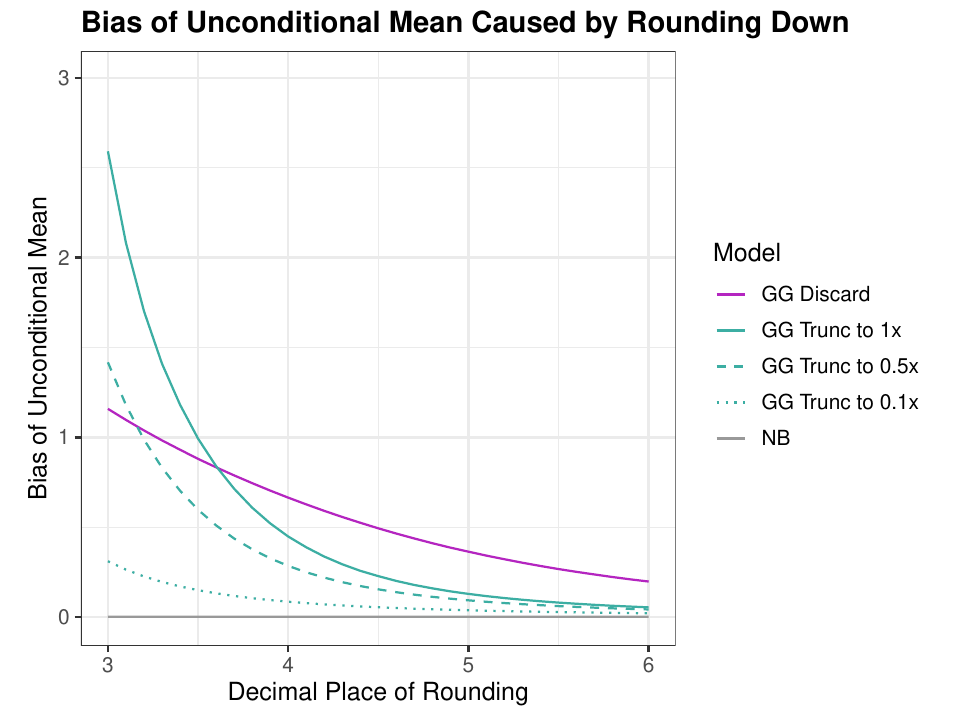}
\caption{The bias of the unconditional mean given by the generalized gamma model (GG) with zeros discarded (Discard) or truncated (Trunc) and  the negative binomial model (NB).}
\label{fig:simRounding}
\end{center}
\end{figure}

Let us conduct the following experiment to explore the influence of rounding on the estimation of GAS models based on discrete and continuous distributions. We simulate 10000 observations using a dynamic model based on the generalized gamma distribution with the time-varying scale parameter following the GAS dynamics given by $c=0.10$, $a=0.10$, $b=0.90$ and the two static shape parameters $\theta=0.50$ and $\varphi=0.50$. The unconditional mean is then approximately equal to 2.05. Then, we round down the observations to a given number of decimal places. Finally, using rounded observations, we estimate GAS models based on the generalized gamma distribution with zero values (created by the rounding) either discarded or truncated as well as the GAS model based on the negative binomial distribution. Note that we do not consider zero inflation in the negative binomial distribution as there are no excessive zeros generated by a different process. The simulation is repeated 1000 times. Figure \ref{fig:simRounding} shows the bias of the unconditional mean of the estimated models with data rounded down to decimal places ranging from 3 up to 6. The negative binomial model, although with incorrectly specified distribution, has the smallest bias. On the other hand, the generalized gamma model with either treatment of zero values has a much higher bias which increases with rounding to fewer decimal places. This is caused by an increased occurrence of discarded or truncated zero values which significantly distorts the continuous distribution. This experiment demonstrates that it is more appropriate to use a distribution that is able to handle zero values, even though it is not the true distribution of the data generating process.

\subsection{Other Mixture Models}
\label{sec:discMix}

On a final note, we discuss some potential alternatives to our proposed model that also utilize a mixture of two processes to capture unrelated and split transactions.

One possibility is to consider a hurdle model based on a continuous distribution with a point mass at zero. For example, the dynamic zero-augmented model of \cite{Hautsch2014}\footnote{The use of zero-augmented models for duration modeling was suggested by Prof.~T.~V.~Ramanathan during the 3rd Conference and Workshop on Statistical Methods in Finance (Chennai, December 16--19, 2017).} or the dynamic censoring model of \cite{Harvey2020} could be used. \cite{Hautsch2014} proposed a multiplicative error model based on a zero-augmented distribution and applied it to high-frequency time series of cumulated trading volumes. \cite{Harvey2020} proposed a dynamic model with a left-shifted distribution for non-zero observations and censored negative values and applied it to daily rainfalls in northern Australia. Note that similarly to us, \cite{Harvey2020} utilized the GAS framework. There are, however, two issues with this approach. Without any transformation of data, both these models would require split transactions to result in exactly zero durations, which is not realistic as shown in Section \ref{sec:intro}. Of course, one could follow our approach and round down durations below a given threshold, e.g.\ one hundredth of a second, to zero. Unlike in our approach, only durations below the threshold would be rounded and durations above would be kept continuous. The second issue is that hurdle models assume that one process generates zero values while the other process generates positive values only. In other words, it would not be possible to determine the ratio between zeros caused by unrelated and split transactions as all zeros would be attributed solely to split transactions. For this reason, our proposed model is superior.

A more complex approach is to assume a non-trivial process for split transactions. Both processes would then generate positive values and at least one of them would also generate zero values. This could be accomplished within either a continuous or discrete framework depending on the underlying data. The choice of a continuous distribution for the process governing split transactions would, however, be limited as zero is required to lie in its support. An exponential distribution would be an obvious starting point here. Note that the appropriately chosen process governing split transactions would not require any transformation of data, which would be a major benefit. On the other hand, the potential complexity of such a model could be a drawback. The ACD model based on a mixture of two non-trivial processes is the direction of our future research.

\section{Conclusion}
\label{sec:conclusion}

We analyze trade durations with split transactions manifesting themselves as zero and close-to-zero values. We round down durations to hundredths of a second and approach this problem within a discrete framework. To capture excessive zero values and autocorrelation structure in durations, we propose a model based on the zero-inflated negative binomial distribution with score dynamics for the time-varying parameters. We label this model the zero-inflated autoregressive conditional duration model or ZIACD model for short. The paper has three main contributions.
\begin{enumerate}
\item We extend the theory of GAS models for the zero-inflated negative binomial distribution with time-varying scale parameter. Specifically, we establish the invertibility of the score filter. We also derive sufficient conditions for the consistency and asymptotic normality of the maximum likelihood of the model parameters.
\item We argue that zero durations should not be removed from the data as they can correspond not only to split transactions but to unrelated transactions as well. Even more, split transactions can generate not only zero values but positive values as well. In the empirical study, the proposed model identifies that split transactions form between 92 and 98 percent of durations smaller than 0.01 seconds. Furthermore, between 53 and 75 percent of all durations correspond to split transactions.
\item We compare the proposed discrete approach with the commonly used continuous approach. We find that even when durations are recorded with high precision suitable for continuous modeling, the proposed discrete model estimated from rounded durations outperforms traditional continuous models based on unrounded data due to its correct treatment of zero and close-to-zero values.
\end{enumerate}
Our proposed model can be utilized in joint modeling of prices and durations. It also allows studying the trading process from the market microstructure perspective. Future research should focus on more complex mixture models, whether in discrete or continuous frameworks, that do not require any transformation of data. However, it should be noted, that these complex models might lose the benefits of our ZIACD model such as simple implementability in practice and verifiability of sufficient conditions for asymptotic properties of the estimator.

\section*{Funding}

The work of Francisco Blasques was supported by the Dutch Science Foundation (NWO) under project VI.Vidi.195.099. The work of Vladimír Holý was supported by the Internal Grant Agency of the University of Economics, Prague under project F4/21/2018. The work of Petra Tomanová was supported by the Czech Science Foundation under project 23-06139S.

\section*{Acknowledgements}
\label{sec:acknow}

Computational resources were supplied by the project ``e-Infrastruktura CZ'' (e-INFRA LM2018140) provided within the program Projects of Large Research, Development and Innovations Infrastructures. We would like to thank Michal Černý and Tomáš Cipra for their comments. We would also like to thank participants of the 61st Meeting of EURO Working Group for Commodities and Financial Modelling (Kaunas, May 16--18, 2018) and the 2nd International Conference on Econometrics and Statistics (Hong Kong, June 19--21, 2018) for fruitful discussions.

%\bibliography{library.bib}

\begin{thebibliography}{85}
\newcommand{\enquote}[1]{``#1''}
\providecommand{\natexlab}[1]{#1}
\providecommand{\url}[1]{\texttt{#1}}
\providecommand{\urlprefix}{}
\expandafter\ifx\csname urlstyle\endcsname\relax
  \providecommand{\doi}[1]{doi:\discretionary{}{}{}#1}\else
  \providecommand{\doi}{doi:\discretionary{}{}{}\begingroup
  \urlstyle{rm}\Url}\fi
\providecommand{\eprint}[2][]{\url{#2}}

\bibitem[{Akaike(1973)}]{Akaike1973}
Akaike H (1973).
\newblock \enquote{{Information Theory and an Extension of the Maximum
  Likelihood Principle}.}
\newblock In \emph{Proceedings of the 2nd International Symposium on
  Information Theory},  267--281. Budapest.
\newblock
  \urlprefix\url{https://link.springer.com/chapter/10.1007/978-1-4612-1694-0{\_}15}.

\bibitem[{Akaike(1974)}]{Akaike1974}
Akaike H (1974).
\newblock \enquote{{A New Look at the Statistical Model Identification}.}
\newblock \emph{IEEE Transactions on Automatic Control}, \textbf{19}(6),
  716--723.
\newblock ISSN 0018-9286.
\newblock \url{https://doi.org/10.1109/tac.1974.1100705}.

\bibitem[{Amisano and Giacomini(2007)}]{Amisano2007}
Amisano G, Giacomini R (2007).
\newblock \enquote{{Comparing Density Forecasts via Weighted Likelihood Ratio
  Tests}.}
\newblock \emph{Journal of Business {\&} Economic Statistics}, \textbf{25}(2),
  177--190.
\newblock ISSN 0735-0015.
\newblock \url{https://doi.org/10.1198/073500106000000332}.

\bibitem[{Andr{\'{e}}e \emph{et~al.}(2017)Andr{\'{e}}e, Blasques, and
  Koomen}]{Andree2017}
Andr{\'{e}}e BPJ, Blasques F, Koomen E (2017).
\newblock \enquote{{Smooth Transition Spatial Autoregressive Models}.}
\newblock \urlprefix\url{https://ssrn.com/abstract=2977830}.

\bibitem[{Andres and Harvey(2012)}]{Andres2012}
Andres P, Harvey A (2012).
\newblock \enquote{{The Dynamic Location/Scale Model}.}
\newblock \url{https://doi.org/10.17863/cam.4972}.

\bibitem[{Bao \emph{et~al.}(2007)Bao, Lee, and Saltoǧlu}]{Bao2007b}
Bao Y, Lee TH, Saltoǧlu B (2007).
\newblock \enquote{{Comparing Density Forecast Models}.}
\newblock \emph{Journal of Forecasting}, \textbf{26}(3), 203--225.
\newblock ISSN 0277-6693.
\newblock \url{https://doi.org/10.1002/for.1023}.

\bibitem[{Bauwens(2006)}]{Bauwens2006}
Bauwens L (2006).
\newblock \enquote{{Econometric Analysis of Intra-Daily Trading Activity on the
  Tokyo Stock Exchange}.}
\newblock \emph{Monetary and Economic Studies}, \textbf{24}(1), 1--24.
\newblock ISSN 0288-8432.
\newblock
  \urlprefix\url{http://www.imes.boj.or.jp/research/abstracts/english/me24-1-1.html}.

\bibitem[{Bauwens and Giot(2000)}]{Bauwens2000}
Bauwens L, Giot P (2000).
\newblock \enquote{{The Logarithmic ACD Model: An Application to the Bid-Ask
  Quote Process of Three NYSE Stocks}.}
\newblock \emph{Annales d'{\'{E}}conomie et de Statistique}, \textbf{60},
  117--149.
\newblock ISSN 0769-489X.
\newblock \url{https://doi.org/10.2307/20076257}.

\bibitem[{Bauwens and Giot(2003)}]{Bauwens2003}
Bauwens L, Giot P (2003).
\newblock \enquote{{Asymmetric ACD Models: Introducing Price Information in ACD
  Models}.}
\newblock \emph{Empirical Economics}, \textbf{28}(4), 709--731.
\newblock ISSN 0377-7332.
\newblock \url{https://doi.org/10.1007/s00181-003-0155-7}.

\bibitem[{Bauwens \emph{et~al.}(2004)Bauwens, Giot, Grammig, and
  Veredas}]{Bauwens2004}
Bauwens L, Giot P, Grammig J, Veredas D (2004).
\newblock \enquote{{A Comparison of Financial Duration Models via Density
  Forecasts}.}
\newblock \emph{International Journal of Forecasting}, \textbf{20}(4),
  589--609.
\newblock ISSN 0169-2070.
\newblock \url{https://doi.org/10.1016/j.ijforecast.2003.09.014}.

\bibitem[{Bauwens and Hautsch(2009)}]{Bauwens2009}
Bauwens L, Hautsch N (2009).
\newblock \enquote{{Modelling Financial High Frequency Data Using Point
  Processes}.}
\newblock In \emph{Handbook of Financial Time Series}, first Edition,
  Chapter~41,  953--979. Springer, Berlin, Heidelberg.
\newblock ISBN 978-3-540-71296-1.
\newblock \url{https://doi.org/10.1007/978-3-540-71297-8}.

\bibitem[{Bauwens and Veredas(2004)}]{Bauwens2004a}
Bauwens L, Veredas D (2004).
\newblock \enquote{{The Stochastic Conditional Duration Model: A Latent
  Variable Model for the Analysis of Financial Durations}.}
\newblock \emph{Journal of Econometrics}, \textbf{119}(2), 381--412.
\newblock ISSN 0304-4076.
\newblock \url{https://doi.org/10.1016/s0304-4076(03)00201-x}.

\bibitem[{Bhatti(2010)}]{Bhatti2010}
Bhatti CR (2010).
\newblock \enquote{{The Birnbaum-Saunders Autoregressive Conditional Duration
  Model}.}
\newblock \emph{Mathematics and Computers in Simulation}, \textbf{80}(10),
  2062--2078.
\newblock ISSN 0378-4754.
\newblock \url{https://doi.org/10.1016/j.matcom.2010.01.011}.

\bibitem[{Blasques \emph{et~al.}(2015)Blasques, Koopman, and
  Lucas}]{Blasques2015}
Blasques F, Koopman SJ, Lucas A (2015).
\newblock \enquote{{Information-Theoretic Optimality of Observation-Driven Time
  Series Models for Continuous Responses}.}
\newblock \emph{Biometrika}, \textbf{102}(2), 325--343.
\newblock ISSN 0006-3444.
\newblock \url{https://doi.org/10.1093/biomet/asu076}.

\bibitem[{Blasques \emph{et~al.}(2022)Blasques, van Brummelen, Koopman, and
  Lucas}]{Blasques2022}
Blasques F, van Brummelen J, Koopman SJ, Lucas A (2022).
\newblock \enquote{{Maximum Likelihood Estimation for Score-Driven Models}.}
\newblock \emph{Journal of Econometrics}, \textbf{227}(2), 325--346.
\newblock ISSN 0304-4076.
\newblock \url{https://doi.org/10.1016/j.jeconom.2021.06.003}.

\bibitem[{Bortoluzzo \emph{et~al.}(2010)Bortoluzzo, Morettin, and
  Toloi}]{Bortoluzzo2010}
Bortoluzzo AB, Morettin PA, Toloi CMC (2010).
\newblock \enquote{{Time-Varying Autoregressive Conditional Duration Model}.}
\newblock \emph{Journal of Applied Statistics}, \textbf{37}(5), 847--864.
\newblock ISSN 0266-4763.
\newblock \url{https://doi.org/10.1080/02664760902914458}.

\bibitem[{Boswell and Patil(1970)}]{Boswell1970}
Boswell M, Patil GP (1970).
\newblock \enquote{{Chance Mechanisms Generating the Negative Binomial
  Distribution}.}
\newblock In GP~Patil (Ed.), \emph{Random Counts in Models and Structures},
  Volume~1,  3--22. Penn State University Press.
\newblock
  \urlprefix\url{http://www.psupress.org/books/titles/0-271-00114-3.html}.

\bibitem[{Bougerol(1993)}]{Bougerol1993}
Bougerol P (1993).
\newblock \enquote{{Kalman Filtering with Random Coefficients and
  Contractions}.}
\newblock \emph{SIAM Journal on Control and Optimization}, \textbf{31}(4),
  942--959.
\newblock ISSN 0363-0129.
\newblock \url{https://doi.org/10.1137/0331041}.

\bibitem[{Cameron and Trivedi(1986)}]{Cameron1986}
Cameron AC, Trivedi PK (1986).
\newblock \enquote{{Econometric Models Based on Count Data: Comparisons and
  Applications of Some Estimators and Tests}.}
\newblock \emph{Journal of Applied Econometrics}, \textbf{1}(1), 29--53.
\newblock ISSN 0883-7252.
\newblock \url{https://doi.org/10.1002/jae.3950010104}.

\bibitem[{Cameron and Trivedi(2013)}]{Cameron2013}
Cameron AC, Trivedi PK (2013).
\newblock \emph{{Regression Analysis of Count Data}}.
\newblock Second Edition. Cambridge University Press, New York.
\newblock ISBN 978-1-107-01416-9.
\newblock \url{https://doi.org/10.1017/cbo9781139013567}.

\bibitem[{Chen \emph{et~al.}(2013)Chen, Diebold, and Schorfheide}]{Chen2013}
Chen F, Diebold FX, Schorfheide F (2013).
\newblock \enquote{{A Markov-Switching Multifractal Inter-Trade Duration Model,
  with Application to US Equities}.}
\newblock \emph{Journal of Econometrics}, \textbf{177}(2), 320--342.
\newblock ISSN 0304-4076.
\newblock \url{https://doi.org/10.1016/j.jeconom.2013.04.016}.

\bibitem[{Christou and Fokianos(2014)}]{Christou2014}
Christou V, Fokianos K (2014).
\newblock \enquote{{Quasi-Likelihood Inference for Negative Binomial Time
  Series Models}.}
\newblock \emph{Journal of Time Series Analysis}, \textbf{35}(1), 55--78.
\newblock ISSN 0143-9782.
\newblock \url{https://doi.org/10.1111/jtsa.12050}.

\bibitem[{Cox(1981)}]{Cox1981}
Cox DR (1981).
\newblock \enquote{{Statistical Analysis of Time Series: Some Recent
  Developments}.}
\newblock \emph{Scandinavian Journal of Statistics}, \textbf{8}(2), 93--108.
\newblock ISSN 0303-6898.
\newblock \url{https://doi.org/10.2307/4615819}.

\bibitem[{Creal \emph{et~al.}(2013)Creal, Koopman, and Lucas}]{Creal2013}
Creal D, Koopman SJ, Lucas A (2013).
\newblock \enquote{{Generalized Autoregressive Score Models with
  Applications}.}
\newblock \emph{Journal of Applied Econometrics}, \textbf{28}(5), 777--795.
\newblock ISSN 0883-7252.
\newblock \url{https://doi.org/10.1002/jae.1279}.

\bibitem[{{De Luca} and Gallo(2004)}]{DeLuca2004}
{De Luca} G, Gallo GM (2004).
\newblock \enquote{{Mixture Processes for Financial Intradaily Durations}.}
\newblock \emph{Studies in Nonlinear Dynamics and Econometrics}, \textbf{8}(2),
  1--18.
\newblock ISSN 1081-1826.
\newblock \url{https://doi.org/10.2202/1558-3708.1223}.

\bibitem[{{De Luca} and Gallo(2009)}]{DeLuca2009}
{De Luca} G, Gallo GM (2009).
\newblock \enquote{{Time-Varying Mixing Weights in Mixture Autoregressive
  Conditional Duration Models}.}
\newblock \emph{Econometric Reviews}, \textbf{28}(1-3), 102--120.
\newblock ISSN 0747-4938.
\newblock \url{https://doi.org/10.1080/07474930802387944}.

\bibitem[{{De Luca} and Zuccolotto(2003)}]{DeLuca2003}
{De Luca} G, Zuccolotto P (2003).
\newblock \enquote{{Finite and Infinite Mixtures for Financial Durations}.}
\newblock \emph{Metron - International Journal of Statistics}, \textbf{61}(3),
  431--455.
\newblock ISSN 00261424.
\newblock \urlprefix\url{https://ideas.repec.org/a/mtn/ancoec/030307.html}.

\bibitem[{Diebold and Mariano(1995)}]{Diebold1995}
Diebold FX, Mariano RS (1995).
\newblock \enquote{{Comparing Predictive Accuracy}.}
\newblock \emph{Journal of Business {\&} Economic Statistics}, \textbf{13}(3),
  253--263.
\newblock ISSN 0735-0015.
\newblock \url{https://doi.org/10.1080/07350015.1995.10524599}.

\bibitem[{Diks \emph{et~al.}(2011)Diks, Panchenko, and van Dijk}]{Diks2011}
Diks C, Panchenko V, van Dijk D (2011).
\newblock \enquote{{Likelihood-Based Scoring Rules for Comparing Density
  Forecasts in Tails}.}
\newblock \emph{Journal of Econometrics}, \textbf{163}(2), 215--230.
\newblock ISSN 0304-4076.
\newblock \url{https://doi.org/10.1016/j.jeconom.2011.04.001}.

\bibitem[{Engle(2000)}]{Engle2000}
Engle RF (2000).
\newblock \enquote{{The Econometrics of Ultra-High-Frequency Data}.}
\newblock \emph{Econometrica}, \textbf{68}(1), 1--22.
\newblock ISSN 0012-9682.
\newblock \url{https://doi.org/10.1111/1468-0262.00091}.

\bibitem[{Engle and Russell(1998)}]{Engle1998}
Engle RF, Russell JR (1998).
\newblock \enquote{{Autoregressive Conditional Duration: A New Model for
  Irregularly Spaced Transaction Data}.}
\newblock \emph{Econometrica}, \textbf{66}(5), 1127--1162.
\newblock ISSN 0012-9682.
\newblock \url{https://doi.org/10.2307/2999632}.

\bibitem[{Feng(2004)}]{Feng2004}
Feng D (2004).
\newblock \enquote{{Stochastic Conditional Duration Models with "Leverage
  Effect" for Financial Transaction Data}.}
\newblock \emph{Journal of Financial Econometrics}, \textbf{2}(3), 390--421.
\newblock ISSN 1479-8409.
\newblock \url{https://doi.org/10.1093/jjfinec/nbh016}.

\bibitem[{Fernandes and Grammig(2005)}]{Fernandes2005}
Fernandes M, Grammig J (2005).
\newblock \enquote{{Nonparametric Specification Tests for Conditional Duration
  Models}.}
\newblock \emph{Journal of Econometrics}, \textbf{127}(1), 35--68.
\newblock ISSN 0304-4076.
\newblock \url{https://doi.org/10.1016/j.jeconom.2004.06.003}.

\bibitem[{Fernandes and Grammig(2006)}]{Fernandes2006}
Fernandes M, Grammig J (2006).
\newblock \enquote{{A Family of Autoregressive Conditional Duration Models}.}
\newblock \emph{Journal of Econometrics}, \textbf{130}(1), 1--23.
\newblock ISSN 0304-4076.
\newblock \url{https://doi.org/10.1016/j.jeconom.2004.08.016}.

\bibitem[{Gallant and White(1988)}]{Gallant1988}
Gallant AR, White H (1988).
\newblock \emph{{A Unified Theory of Estimation and Inference for Nonlinear
  Dynamic Models}}.
\newblock First Edition. Basil Blackwell, Oxford.
\newblock ISBN 978-0-631-15765-6.
\newblock \urlprefix\url{https://books.google.com/books?id=VVOqQgAACAAJ}.

\bibitem[{Ghysels \emph{et~al.}(2004)Ghysels, Gouri{\'{e}}roux, and
  Jasiak}]{Ghysels2004a}
Ghysels E, Gouri{\'{e}}roux C, Jasiak J (2004).
\newblock \enquote{{Stochastic Volatility Duration Models}.}
\newblock \emph{Journal of Econometrics}, \textbf{119}(2), 413--433.
\newblock ISSN 0304-4076.
\newblock \url{https://doi.org/10.1016/S0304-4076(03)00202-1}.

\bibitem[{G{\'{o}}mez-D{\'{e}}niz and
  P{\'{e}}rez-Rodr{\'{i}}guez(2016)}]{Gomez-Deniz2016a}
G{\'{o}}mez-D{\'{e}}niz E, P{\'{e}}rez-Rodr{\'{i}}guez JV (2016).
\newblock \enquote{{Conditional Duration Model and the Unobserved Market
  Heterogeneity of Traders: An Infinite Mixture of Non-Exponentials}.}
\newblock \emph{Revista Colombiana de Estadistica}, \textbf{39}(2), 307--323.
\newblock ISSN 01201751.
\newblock \url{https://doi.org/10.15446/rce.v39n2.51584}.

\bibitem[{G{\'{o}}mez-D{\'{e}}niz and
  P{\'{e}}rez-Rodr{\'{i}}guez(2017)}]{Gomez-Deniz2017}
G{\'{o}}mez-D{\'{e}}niz E, P{\'{e}}rez-Rodr{\'{i}}guez JV (2017).
\newblock \enquote{{Mixture Inverse Gaussian for Unobserved Heterogeneity in
  the Autoregressive Conditional Duration Model}.}
\newblock \emph{Communications in Statistics - Theory and Methods},
  \textbf{46}(18), 9007--9025.
\newblock ISSN 0361-0926.
\newblock \url{https://doi.org/10.1080/03610926.2016.1200094}.

\bibitem[{Gorgi(2018)}]{Gorgi2018}
Gorgi P (2018).
\newblock \enquote{{Integer-Valued Autoregressive Models with Survival
  Probability Driven By a Stochastic Recurrence Equation}.}
\newblock \emph{Journal of Time Series Analysis}, \textbf{39}(2), 150--171.
\newblock ISSN 0143-9782.
\newblock \url{https://doi.org/10.1111/jtsa.12272}.

\bibitem[{Grammig and Maurer(2000)}]{Grammig2000}
Grammig J, Maurer KO (2000).
\newblock \enquote{{Non-Monotonic Hazard Functions and the Autoregressive
  Conditional Duration Model}.}
\newblock \emph{The Econometrics Journal}, \textbf{3}(1), 16--38.
\newblock ISSN 1368-4221.
\newblock \url{https://doi.org/10.1111/1368-423x.00037}.

\bibitem[{Grammig and Wellner(2002)}]{Grammig2002}
Grammig J, Wellner M (2002).
\newblock \enquote{{Modeling the Interdependence of Volatility and
  Inter-Transaction Duration Processes}.}
\newblock \emph{Journal of Econometrics}, \textbf{106}(2), 369--400.
\newblock \url{https://doi.org/10.1016/S0304-4076(01)00105-1}.

\bibitem[{Greene(1994)}]{Greene1994}
Greene WH (1994).
\newblock \enquote{{Accounting for Excess Zeros and Sample Selection in Poisson
  and Negative Binomial Regression Models}.}
\newblock \urlprefix\url{http://ssrn.com/abstract=1293115}.

\bibitem[{Grimshaw \emph{et~al.}(2005)Grimshaw, McDonald, McQueen, and
  Thorley}]{Grimshaw2005}
Grimshaw SD, McDonald J, McQueen GR, Thorley S (2005).
\newblock \enquote{{Estimating Hazard Functions for Discrete Lifetimes}.}
\newblock \emph{Communications in Statistics - Simulation and Computation},
  \textbf{34}(2), 451--463.
\newblock ISSN 0361-0918.
\newblock \url{https://doi.org/10.1081/SAC-200055732}.

\bibitem[{Harvey(2013)}]{Harvey2013}
Harvey AC (2013).
\newblock \emph{{Dynamic Models for Volatility and Heavy Tails: With
  Applications to Financial and Economic Time Series}}.
\newblock First Edition. Cambridge University Press, New York.
\newblock ISBN 978-1-107-63002-4.
\newblock \url{https://doi.org/10.1017/cbo9781139540933}.

\bibitem[{Harvey and Ito(2020)}]{Harvey2020}
Harvey AC, Ito R (2020).
\newblock \enquote{{Modeling Time Series When Some Observations Are Zero}.}
\newblock \emph{Journal of Econometrics}, \textbf{214}(1), 33--45.
\newblock ISSN 0304-4076.
\newblock \url{https://doi.org/10.1016/j.jeconom.2019.05.003}.

\bibitem[{Hautsch(2001)}]{Hautsch2001}
Hautsch N (2001).
\newblock \enquote{{Modelling Intraday Trading Activity Using Box-Cox ACD
  Models}.}
\newblock \url{https://doi.org/10.2139/ssrn.289643}.
\newblock \urlprefix\url{https://ssrn.com/abstract=289643}.

\bibitem[{Hautsch(2003)}]{Hautsch2003}
Hautsch N (2003).
\newblock \enquote{{Assessing the Risk of Liquidity Suppliers on the Basis of
  Excess Demand Intensities}.}
\newblock \emph{Journal of Financial Econometrics}, \textbf{1}(2), 189--215.
\newblock ISSN 1479-8409.
\newblock \url{https://doi.org/10.1093/jjfinec/nbg010}.

\bibitem[{Hautsch(2012)}]{Hautsch2011}
Hautsch N (2012).
\newblock \emph{{Econometrics of Financial High-Frequency Data}}.
\newblock First Edition. Springer, Berlin, Heidelberg.
\newblock ISBN 978-3-642-21924-5.
\newblock \url{https://doi.org/10.1007/978-3-642-21925-2}.

\bibitem[{Hautsch \emph{et~al.}(2014)Hautsch, Malec, and
  Schienle}]{Hautsch2014}
Hautsch N, Malec P, Schienle M (2014).
\newblock \enquote{{Capturing the Zero: A New Class of Zero-Augmented
  Distributions and Multiplicative Error Processes}.}
\newblock \emph{Journal of Financial Econometrics}, \textbf{12}(1), 89--121.
\newblock ISSN 1479-8409.
\newblock \url{https://doi.org/10.1093/jjfinec/nbt002}.

\bibitem[{Herrera and Schipp(2013)}]{Herrera2013}
Herrera R, Schipp B (2013).
\newblock \enquote{{Value at Risk Forecasts by Extreme Value Models in a
  Conditional Duration Framework}.}
\newblock \emph{Journal of Empirical Finance}, \textbf{23}, 33--47.
\newblock ISSN 0927-5398.
\newblock \url{https://doi.org/10.1016/j.jempfin.2013.05.002}.

\bibitem[{Hol{\'{y}}(2020)}]{Holy2020}
Hol{\'{y}} V (2020).
\newblock \enquote{{Impact of the Parametrization and the Scaling Function in
  Dynamic Score-Driven Models: The Case of the Negative Binomial
  Distribution}.}
\newblock In \emph{Proceedings of the 38th International Conference
  Mathematical Methods in Economics},  173--179. Mendel University in Brno,
  Brno.
\newblock ISBN 978-80-7509-734-7.
\newblock
  \urlprefix\url{https://mme2020.mendelu.cz/wcd/w-rek-mme/mme2020{\_}conference{\_}proceedings{\_}final.pdf}.

\bibitem[{Hol{\'{y}} and Tomanov{\'{a}}(2022)}]{Holy2022b}
Hol{\'{y}} V, Tomanov{\'{a}} P (2022).
\newblock \enquote{{Modeling Price Clustering in High-Frequency Prices}.}
\newblock \emph{Quantitative Finance}, \textbf{22}(9), 1649--1663.
\newblock ISSN 1469-7688.
\newblock \url{https://doi.org/10.1080/14697688.2022.2050285}.

\bibitem[{Hujer \emph{et~al.}(2005)Hujer, Vuletic, and Kokot}]{Hujer2005}
Hujer R, Vuletic S, Kokot S (2005).
\newblock \enquote{{The Markov Switching ACD Model}.}
\newblock \url{https://doi.org/10.2139/ssrn.332381}.

\bibitem[{Jasiak(1998)}]{Jasiak1998}
Jasiak J (1998).
\newblock \enquote{{Persistence in Intertrade Durations}.}
\newblock \emph{Finance}, \textbf{19}, 166--195.
\newblock ISSN 1556-5068.
\newblock \url{https://doi.org/10.2139/ssrn.162008}.

\bibitem[{Jeyasreedharan \emph{et~al.}(2014)Jeyasreedharan, Allen, and
  Yang}]{Jeyasreedharan2014}
Jeyasreedharan N, Allen DE, Yang JW (2014).
\newblock \enquote{{Yet Another ACD Model: The Autoregressive Conditional
  Directional Duration (ACDD) Model}.}
\newblock \emph{Annals of Financial Economics}, \textbf{9}(1),
  1450004/1--1450004/20.
\newblock ISSN 2010-4952.
\newblock \url{https://doi.org/10.1142/S2010495214500043}.

\bibitem[{Konishi and Kitagawa(2008)}]{Konishi2008}
Konishi S, Kitagawa G (2008).
\newblock \emph{{Information Criteria and Statistical Modeling}}.
\newblock Springer Series in Statistics. Springer, New York.
\newblock ISBN 978-0-387-71886-6.
\newblock \url{https://doi.org/10.1007/978-0-387-71887-3}.

\bibitem[{Koopman and Lit(2019)}]{Koopman2019}
Koopman SJ, Lit R (2019).
\newblock \enquote{{Forecasting Football Match Results in National League
  Competitions Using Score-Driven Time Series Models}.}
\newblock \emph{International Journal of Forecasting}, \textbf{35}(2),
  797--809.
\newblock ISSN 0169-2070.
\newblock \url{https://doi.org/10.1016/j.ijforecast.2018.10.011}.

\bibitem[{Koopman \emph{et~al.}(2018)Koopman, Lit, Lucas, and
  Opschoor}]{Koopman2018}
Koopman SJ, Lit R, Lucas A, Opschoor A (2018).
\newblock \enquote{{Dynamic Discrete Copula Models for High-Frequency Stock
  Price Changes}.}
\newblock \emph{Journal of Applied Econometrics}, \textbf{33}(7), 966--985.
\newblock ISSN 0883-7252.
\newblock \url{https://doi.org/10.1002/jae.2645}.

\bibitem[{Koopman \emph{et~al.}(2016)Koopman, Lucas, and Scharth}]{Koopman2016}
Koopman SJ, Lucas A, Scharth M (2016).
\newblock \enquote{{Predicting Time-Varying Parameters with Parameter-Driven
  and Observation-Driven Models}.}
\newblock \emph{Review of Economics and Statistics}, \textbf{98}(1), 97--110.
\newblock ISSN 0034-6535.
\newblock \url{https://doi.org/10.1162/rest_a_00533}.

\bibitem[{Lambert(1992)}]{Lambert1992}
Lambert D (1992).
\newblock \enquote{{Zero-Inflated Poisson Regression, with an Application to
  Defects in Manufacturing}.}
\newblock \emph{Technometrics}, \textbf{34}(1), 1--14.
\newblock ISSN 0040-1706.
\newblock \url{https://doi.org/10.2307/1269547}.

\bibitem[{Leiva \emph{et~al.}(2014)Leiva, Saulo, Le{\~{a}}o, and
  Marchant}]{Leiva2014}
Leiva V, Saulo H, Le{\~{a}}o J, Marchant C (2014).
\newblock \enquote{{A Family of Autoregressive Conditional Duration Models
  Applied to Financial Data}.}
\newblock \emph{Computational Statistics {\&} Data Analysis}, \textbf{79},
  175--191.
\newblock ISSN 0167-9473.
\newblock \url{https://doi.org/10.1016/j.csda.2014.05.016}.

\bibitem[{Li and Bai(2011)}]{Li2011}
Li W, Bai ZD (2011).
\newblock \enquote{{Analysis of Accumulated Rounding Errors in Autoregressive
  Processes}.}
\newblock \emph{Journal of Time Series Analysis}, \textbf{32}(5), 518--530.
\newblock ISSN 0143-9782.
\newblock \url{https://doi.org/10.1111/j.1467-9892.2010.00710.x}.

\bibitem[{Liu \emph{et~al.}(2018)Liu, Kong, and Jing}]{Liu2018}
Liu Z, Kong XB, Jing BY (2018).
\newblock \enquote{{Estimating the Integrated Volatility Using High-Frequency
  Data with Zero Durations}.}
\newblock \emph{Journal of Econometrics}, \textbf{204}(1), 18--32.
\newblock ISSN 0304-4076.
\newblock \url{https://doi.org/10.1016/j.jeconom.2017.12.008}.

\bibitem[{Lunde(1999)}]{Lunde1999a}
Lunde A (1999).
\newblock \enquote{{A Generalized Gamma Autoregressive Conditional Duration
  Model}.}
\newblock \urlprefix\url{https://www.researchgate.net/publication/228464216}.

\bibitem[{Mishra and Ramanathan(2017)}]{Mishra2017}
Mishra A, Ramanathan TV (2017).
\newblock \enquote{{Nonstationary Autoregressive Conditional Duration Models}.}
\newblock \emph{Studies in Nonlinear Dynamics and Econometrics},
  \textbf{21}(4), 1--22.
\newblock ISSN 1081-1826.
\newblock \url{https://doi.org/10.1515/snde-2015-0057}.

\bibitem[{Pacurar(2008)}]{Pacurar2008}
Pacurar M (2008).
\newblock \enquote{{Autoregressive Conditional Duration Models in Finance: A
  Survey of the Theoretical and Empirical Literature}.}
\newblock \emph{Journal of Economic Surveys}, \textbf{22}(4), 711--751.
\newblock ISSN 0950-0804.
\newblock \url{https://doi.org/10.1111/j.1467-6419.2007.00547.x}.

\bibitem[{Rao(1962)}]{Rao1962}
Rao RR (1962).
\newblock \enquote{{Relations between Weak and Uniform Convergence of Measures
  with Applications}.}
\newblock \emph{The Annals of Mathematical Statistics}, \textbf{33}(2),
  659--680.
\newblock ISSN 0003-4851.
\newblock \url{https://doi.org/10.2307/2237541}.

\bibitem[{Russell and Engle(2005)}]{Russell2005}
Russell JR, Engle RF (2005).
\newblock \enquote{{A Discrete-State Continuous-Time Model of Financial
  Transactions Prices and Times: The Autoregressive Conditional
  Multinomial-Autoregressive Conditional Duration Model}.}
\newblock \emph{Journal of Business {\&} Economic Statistics}, \textbf{23}(2),
  166--180.
\newblock ISSN 0735-0015.
\newblock \url{https://doi.org/10.1198/073500104000000541}.

\bibitem[{Saranjeet and Ramanathan(2018)}]{Saranjeet2019}
Saranjeet KB, Ramanathan TV (2018).
\newblock \enquote{{Conditional Duration Models for High-Frequency Data: A
  Review on Recent Developments}.}
\newblock \emph{Journal of Economic Surveys}, \textbf{33}(1), 252--273.
\newblock ISSN 0950-0804.
\newblock \url{https://doi.org/10.1111/joes.12261}.

\bibitem[{Schneeweiss \emph{et~al.}(2010)Schneeweiss, Komlos, and
  Ahmad}]{Schneeweiss2010}
Schneeweiss H, Komlos J, Ahmad AS (2010).
\newblock \enquote{{Symmetric and Asymmetric Rounding: A Review and Some New
  Results}.}
\newblock \emph{AStA Advances in Statistical Analysis}, \textbf{94}(3),
  247--271.
\newblock ISSN 1863-8171.
\newblock \url{https://doi.org/10.1007/s10182-010-0125-2}.

\bibitem[{Sin and White(1996)}]{Sin1996}
Sin CY, White H (1996).
\newblock \enquote{{Information Criteria for Selecting Possibly Misspecified
  Parametric Models}.}
\newblock \emph{Journal of Econometrics}, \textbf{71}(1-2), 207--225.
\newblock ISSN 0304-4076.
\newblock \url{https://doi.org/10.1016/0304-4076(94)01701-8}.

\bibitem[{Stacy(1962)}]{Stacy1962}
Stacy EW (1962).
\newblock \enquote{{A Generalization of the Gamma Distribution}.}
\newblock \emph{The Annals of Mathematical Statistics}, \textbf{33}(3),
  1187--1192.
\newblock ISSN 0003-4851.
\newblock \url{https://doi.org/10.2307/2237889}.

\bibitem[{Straumann and Mikosch(2006)}]{Straumann2006}
Straumann D, Mikosch T (2006).
\newblock \enquote{{Quasi-Maximum-Likelihood Estimation in Conditionally
  Heteroscedastic Time Series: A Stochastic Recurrence Equations Approach}.}
\newblock \emph{The Annals of Statistics}, \textbf{34}(5), 2449--2495.
\newblock ISSN 0090-5364.
\newblock \url{https://doi.org/10.1214/009053606000000803}.

\bibitem[{Taraldsen(2011)}]{Taraldsen2011}
Taraldsen G (2011).
\newblock \enquote{{Analysis of Rounded Exponential Data}.}
\newblock \emph{Journal of Applied Statistics}, \textbf{38}(5), 977--986.
\newblock ISSN 0266-4763.
\newblock \url{https://doi.org/10.1080/02664761003692431}.

\bibitem[{Tomanov{\'{a}} and Hol{\'{y}}(2021)}]{Tomanova2021}
Tomanov{\'{a}} P, Hol{\'{y}} V (2021).
\newblock \enquote{{Clustering of Arrivals in Queueing Systems: Autoregressive
  Conditional Duration Approach}.}
\newblock \emph{Central European Journal of Operations Research},
  \textbf{29}(3), 859--874.
\newblock ISSN 1435-246X.
\newblock \url{https://doi.org/10.1007/s10100-021-00744-7}.

\bibitem[{Tricker(1992)}]{Tricker1992}
Tricker AR (1992).
\newblock \enquote{{Estimation of Parameters for Rounded Data from Non-Normal
  Distributions}.}
\newblock \emph{Journal of Applied Statistics}, \textbf{19}(4), 465--471.
\newblock ISSN 0266-4763.
\newblock \url{https://doi.org/10.1080/02664769200000041}.

\bibitem[{Tricker(1984)}]{Tricker1984a}
Tricker T (1984).
\newblock \enquote{{Effects of Rounding Data Sampled from the Exponential
  Distribution}.}
\newblock \emph{Journal of Applied Statistics}, \textbf{11}(1), 54--87.
\newblock ISSN 0266-4763.
\newblock \url{https://doi.org/10.1080/02664768400000007}.

\bibitem[{Veredas \emph{et~al.}(2002)Veredas, Rodr{\'{i}}guez-Poo, and
  Espasa}]{Veredas2002}
Veredas D, Rodr{\'{i}}guez-Poo JM, Espasa A (2002).
\newblock \enquote{{On the (Intradaily) Seasonality and Dynamics of a Financial
  Point Process: A Semiparametric Approach.}}
\newblock \urlprefix\url{https://ideas.repec.org/p/cor/louvco/2002023.html}.

\bibitem[{White(1994)}]{White1994}
White H (1994).
\newblock \emph{{Estimation, Inference and Specification Analysis}}.
\newblock First Edition. Cambridge University Press, Cambridge.
\newblock ISBN 978-0-521-57446-4.
\newblock \url{https://doi.org/10.1017/CCOL0521252806}.

\bibitem[{Wintenberger(2013)}]{Wintenberger2013}
Wintenberger O (2013).
\newblock \enquote{{Continuous Invertibility and Stable QML Estimation of the
  EGARCH(1,1) Model}.}
\newblock \emph{Scandinavian Journal of Statistics}, \textbf{40}(4), 846--867.
\newblock ISSN 0303-6898.
\newblock \url{https://doi.org/10.1111/sjos.12038}.

\bibitem[{Xu \emph{et~al.}(2011)Xu, Knight, and Wirjanto}]{Xu2011}
Xu D, Knight J, Wirjanto TS (2011).
\newblock \enquote{{Asymmetric Stochastic Conditional Duration Model - A
  Mixture-of-Normal Approach}.}
\newblock \emph{Journal of Financial Econometrics}, \textbf{9}(3), 469--488.
\newblock ISSN 1479-8409.
\newblock \url{https://doi.org/10.1093/jjfinec/nbq026}.

\bibitem[{Xu(2013)}]{Xu2013}
Xu Y (2013).
\newblock \enquote{{The Lognormal Autoregressive Conditional Duration (LNACD)
  Model and a Comparison with an Alternative ACD Models}.}
\newblock \urlprefix\url{https://ssrn.com/abstract=2382159}.

\bibitem[{Zhang \emph{et~al.}(2010)Zhang, Liu, and Bai}]{Zhang2010}
Zhang B, Liu T, Bai ZD (2010).
\newblock \enquote{{Analysis of Rounded Data from Dependent Sequences}.}
\newblock \emph{Annals of the Institute of Statistical Mathematics},
  \textbf{62}(6), 1143--1173.
\newblock ISSN 0020-3157.
\newblock \url{https://doi.org/10.1007/s10463-009-0224-6}.

\bibitem[{Zhang \emph{et~al.}(2001)Zhang, Russell, and Tsay}]{Zhang2001}
Zhang MY, Russell JR, Tsay RS (2001).
\newblock \enquote{{A Nonlinear Autoregressive Conditional Duration Model with
  Applications to Financial Transaction Data}.}
\newblock \emph{Journal of Econometrics}, \textbf{104}(1), 179--207.
\newblock ISSN 0304-4076.
\newblock \url{https://doi.org/10.1016/s0304-4076(01)00063-x}.

\bibitem[{Zheng \emph{et~al.}(2016)Zheng, Li, and Li}]{Zheng2016}
Zheng Y, Li Y, Li G (2016).
\newblock \enquote{{On Fr{\'{e}}chet Autoregressive Conditional Duration
  Models}.}
\newblock \emph{Journal of Statistical Planning and Inference}, \textbf{175},
  51--66.
\newblock ISSN 0378-3758.
\newblock \url{https://doi.org/10.1016/j.jspi.2016.02.009}.

\end{thebibliography}
%\bibliographystyle{myjss}

\appendix

\section{Proofs of Asymptotic Properties}
\label{app:proofs}

\noindent \textit{Proof of Proposition \ref{prop:invertibility}:} 

Following \cite{Straumann2006} and \cite{Blasques2022}, we obtain invertibility by verifying that the conditions of Theorem 3.1 of \cite{Bougerol1993} hold uniformly on a non-empty set $\Theta$, for any initialization $\hat{f}_{1}(\theta)$
 In particular, we note that a $\ln^{+}$ bounded moment holds at $i=1$ since 
\begin{equation*}
\begin{aligned}
\mathrm{E} \left[ \log^{+} \sup_{\theta \in \Theta} \left| c + b \hat{f}_{1}(\theta) + a s(x_{1},\hat{f}_{1}(\theta)) \right| \right] &\leq 4 \ln 2 + \mathrm{E} \left[ \ln^{+} \sup_{\theta \in \Theta}|c| \right] + \mathrm{E} \left[ \ln^{+} \sup_{\theta \in \Theta} \left| b \hat{f}_{1}(\theta) \right| \right] \\
& \quad + \mathrm{E} \left[ \ln^{+} \sup_{\theta \in \Theta} \left| a s(x_{1},\hat{f}_{1}(\theta)) \right| \right] \\
&\leq 4 \ln 2 +  \mathrm{E} \left[ \ln^{+} \sup_{\theta \in \Theta}|c| \right] + \mathrm{E} \left[ \ln^{+} \sup_{\theta \in \Theta}|b| \right] \\
& \quad + \mathrm{E} \left[ \ln^{+} \sup_{\theta \in \Theta} \left| \hat{f}_{1}(\theta) \right| \right] + \mathrm{E} \left[ \ln^{+} \sup_{\theta \in \Theta}|a| \right] \\
& \quad + \mathrm{E} \left[   \sup_{\theta \in \Theta} \left| s(x_{1},\hat{f}_{1}(\theta)) \right| \right] \\
&\leq 4 \ln 2 +      \max\{|c^{-}|,|c^{+}|\}  +     \max\{|b^{-}|,|b^{+}|\} \\
& \quad  +    \sup_{\theta \in \Theta} \left| \hat{f}_{1}(\theta) \right|   +      \max\{|a^{-}|,|a^{+}|\}  + \mathrm{E} \left[ \ln^{+} \sup_{\theta \in \Theta} \left| s(x_{1},\hat{f}_{1}(\theta)) \right| \right] \\
&< \infty, \\
\end{aligned}
\end{equation*} 
where the three inequalities follow by norm sub-additivity, as well as the $\ln^{+}$ sub-additive and sub-multiplicative inequalities in Lemma 2.2 of \cite{Straumann2006}, and the last bound follows since $c$, $b$, $a$ are strictly positive and lie on the compact $\Theta$ and $\hat{f}_{1}(\theta)$ is a given real number. We also have that  $\mathrm{E} \left[ \ln^{+} \sup_{\theta \in \Theta}| s(x_{1},\hat{f}_{1}(\theta))| \right] < \infty$ as
\begin{equation*}
\begin{aligned}
\mathrm{E} \left[ \ln^{+} \sup_{\theta \in \Theta} \left| s(x_{i},\hat{f}_{1}(\theta),\theta) \right| \right] &= \mathrm{P}[x_{i}=0] \cdot \ln^{+}  \sup_{\theta \in \Theta} \left| s(0,\hat f_{1}(\theta),\theta) \right| \\
& \quad + \mathrm{P}[x_{i}>0] \cdot \mathrm{E}_{x_{i}>0} \left[ \ln^{+} \sup_{\theta \in \Theta} \left| s(x_{i},\hat f_{1}(\theta),\theta) \right| \right] \\
&\leq \ \ln^{+}  \sup_{\theta \in \Theta} \left| s(0,\hat f_{1}(\theta),\theta) \right|  + \mathrm{E}_{x_{i}>0} \left[ \ln^{+} \sup_{\theta \in \Theta}  \left| s(x_{i},\hat f_{1}(\theta),\theta) \right| \right] \\
&< \infty, \\
\end{aligned}
\end{equation*}
where $\mathrm{E}_{x_{i}>0}$ denotes the conditional expectation  $\mathrm{E}_{x_{i}>0}[\cdot]=\mathrm{E}[\cdot|x_{i>0}]$ and
\begin{equation*}
\begin{aligned}
\mathrm{E} \left[ \ln^{+} \sup_{\theta \in \Theta}| s(0,\hat{f}_{1},\theta)| \right] &= \ln^{+} \sup_{\theta \in \Theta}| s(0,\hat{f}_{1},\theta)|\\
&= \ln^{+} \sup_{\theta \in \Theta} \Big|	(\pi - 1)   \exp(\hat{f}_1)  (\alpha \exp(\hat{f}_{1}) + 1)^{-1} \\
& \qquad  \cdot \left( 1 + \pi (\alpha \exp(\hat{f}_1) + 1)^{\alpha^{-1}} - \pi \right)^{-1} \Big| \\
&\leq \ln^{+} \sup_{\theta \in \Theta}|\pi - 1| +  \ln^{+} \sup_{\theta \in \Theta}| \exp(\hat{f}_1) | +  \ln^{+} \sup_{\theta \in \Theta}| (\alpha \exp(\hat{f}_{1}) + 1)^{-1}| \\
& \qquad  + \ln^{+} \sup_{\theta \in \Theta} \Big|\Big( 1 + \pi (\alpha \exp(\hat{f}_1) + 1)^{\alpha^{-1}} - \pi \Big)^{-1} \Big| \\
&< \infty,
\end{aligned}
\end{equation*}
which holds as the parameter vector $\theta$ lies on the compact set $\Theta$, and $\hat{f}_{1}$ is a given point in $\mathbb{R}$, and
\begin{equation*}
\begin{aligned}
\mathrm{E}_{x_{i}>0} \left[ \ln^{+} \sup_{\theta \in \Theta} \left| s(x_{i},\hat{f}_{1},\theta) \right| \right] &= \mathrm{E}_{x_{1}>0} \left[ \ln^{+} \sup_{\theta \in \Theta} \left|  x_1 - \exp(\hat{f}_1)(\alpha \exp(\hat{f}_{1}) + 1)^{-1} \right| \right] \\
&\leq \mathrm{E}_{x_{1}>0} \left[ \ln^{+} \sup_{\theta \in \Theta} \left| x_1 - \exp(\hat{f}_1) \right| \right] \\ 
&\leq 2 \ln (2) + \mathrm{E}_{x_{1}>0} \left[ \ln^{+}  |x_1| \right] +  \ln^{+}  | \exp(\hat{f}_1)| \\
&< \infty, \\
\end{aligned}
\end{equation*}
since $x_{1}$ has a logarithmic moment, $\Theta$ is compact and $\hat{f}_{1} \in \mathbb{R}$.

Finally, the contraction condition of \cite{Bougerol1993} is satisfied uniformly in $\theta \in \Theta$ since
\begin{equation*}
\begin{aligned}
&\mathrm{E} \left[ \ln \sup_{ f} \sup_{\theta \in \Theta} \left| a \frac{\partial {s}(x_{i}, f,\theta)}{\partial  f} + b \right| \right] < 0 \\
& \quad \Leftrightarrow \ \mathrm{P}[x_{i}=0] \cdot \ln \sup_{  f} \sup_{\theta \in \Theta} \left| a \frac{\partial {s}(0, f,\theta)}{\partial  f} + b \right| \\
& \qquad + \mathrm{P}[x_{i}>0] \cdot \mathrm{E}_{x_{i}>0} \left[ \ln \sup_{f} \sup_{\theta \in \Theta} \left| a \frac{\partial {s}(x_{i}, f,\theta)}{\partial  f} + b \right| \right] < 0
\end{aligned}
\end{equation*}
where
\begin{equation*}
\begin{aligned}
& \mathrm{E} \left[ \ln \sup_{\hat{f}} \sup_{\theta \in \Theta} \left| a \frac{\partial s(x_{i},\hat{f},\theta)}{\partial \hat{f}} + b \right| \right] < 0 \\
& \quad \Leftrightarrow \ \mathrm{P}[x_{i}=0] \cdot \ln \sup_{\hat{f}} \sup_{\theta \in \Theta} \left| a \frac{\partial s(0,\hat{f},\theta)}{\partial \hat{f}} + b \right| \\
& \qquad + \mathrm{P}[x_{i}>0] \cdot \mathrm{E}_{x_{i}>0} \left[ \ln \sup_{\hat{f}} \sup_{\theta \in \Theta} \left| a \frac{\partial s(x_{i},\hat{f},\theta)}{\partial \hat{f}} + b \right| \right] < 0\\	
& \quad \Leftrightarrow \ \left(\pi + (1 - \pi)  \left( \frac{\alpha^{-1}}{\alpha^{-1} + \hat{f}_i} \right)^{\alpha^{-1}} \right) \\
& \qquad \cdot  \ln \sup_{\hat{f}} \sup_{\theta \in \Theta}  \Bigg|-a \frac{(\pi-1)^{2}\exp(2\hat{f})}{(\alpha \exp(\hat{f})+1)^{2}\left( \pi(\alpha\exp(\hat{f})+1)^{1/\alpha}-\pi+1 \right)^{2}}\\
& \qquad - a\frac{(\pi-1)\exp(\hat{f})(\exp(\hat{f})-1)}{(\alpha \exp(\hat{f})+1)^{2} \left( \pi(\alpha\exp(\hat{f})+1)^{1/\alpha}-\pi+1 \right) }  + b \Bigg| \\ 
&  \qquad + \left( 1-\pi - (1 - \pi)  \left( \frac{\alpha^{-1}}{\alpha^{-1} + \hat{f}_i} \right)^{\alpha^{-1}} \right) \cdot \mathrm{E}_{x_{i}>0} \left[ \ln \sup_{\hat{f}} \sup_{\theta \in \Theta} \left| -a \frac{(\alpha x_{i}+1)\exp(\hat{f})}{(\alpha\exp(\hat{f})+1)^{2}}   + b \right| \right] < 0\\
& \quad \Leftarrow \  \ln \left[  \sup_{\theta \in \Theta} \left| a  \frac{(\pi-1)^{2}}{2 \alpha} \right| + \sup_{\theta \in \Theta} \left| a\frac{(\pi-1)}{\alpha^{2}} \right| + \sup_{\theta \in \Theta}|b| \right] + \mathrm{E}_{x_{i}>0} \left[ \ln \left(   \sup_{\theta \in \Theta} \left| a  \frac{\alpha x_{i}+1 }{4\alpha} \right|  +   \sup_{\theta \in \Theta}  |b| \right) \right]  < 0.\\
\end{aligned}
\end{equation*}
This can be simplified by noting that 			
\begin{equation*}
\begin{aligned}
\frac{\exp(2\hat{f})}{(\alpha \exp(\hat{f})+1)^{2}} &\leq \frac{1}{2 \alpha}, \\
\Big(\pi(\alpha\exp(\hat{f})+1)^{1/\alpha}-\pi+1\Big)^{2} & \geq 1, \\
\frac{\exp(\hat{f})(\exp(\hat{f})-1)}{(\alpha \exp(\hat{f})+1)^{2}} &\leq \frac{1}{\alpha^{2}}. \\
\end{aligned}
\end{equation*}
This, in turn, implies that
\begin{equation*}
\begin{aligned}				
& \mathrm{E} \left[ \ln  \sup_{\hat{f}} \sup_{\theta \in \Theta} \left| a \frac{\partial s(x_{i},\hat{f},\theta)}{\partial \hat{f}} + b \right| \right] < 0 \\			
& \quad \Leftarrow \   \left\{ \ \sup_{\theta \in \Theta}   a  \frac{(\pi-1)^{2}}{2 \alpha}  +   \sup_{\theta \in \Theta} a\frac{|\pi-1|}{\alpha^{2}}   +   \sup_{\theta \in \Theta}b^{+} <1 \  \wedge \ \mathrm{E}_{x_{i}>0} \left[ \ln \left(  \sup_{\theta \in \Theta} a \frac{\alpha x_{i}+1 }{4\alpha}   +   b^{+} \right) \right]  < 0  \ \right\} \\
& \quad \Leftarrow \  \left\{ \  \frac{a^{+}(\pi^{-}-1)^{2}}{2 \alpha^{-}}  +   \frac{a^{+}|\pi^{-}-1|}{(\alpha^{-})^{2}}   +   b^{+} <1 \  \wedge \     \mathrm{E}_{x_{i}>0} \left[ \ln \left( \frac{a^{+} (\alpha^{+} x_{i}+1) }{4\alpha^{-}}   +  b^{+} \right) \right]  < 0  \ \right\}. 
\end{aligned}
\end{equation*}

\medskip

\noindent \textit{Proof of Lemma \ref{theo:consistency}:} 

This proof follows that of  \citet[Theorem 4.6]{Blasques2022}. The  existence and measurability of $\hat{\theta}_{n}$ is obtained through an application of \citet[Theorem 2.11]{White1994} or \citet[Lemma 2.1, Theorem 2.2]{Gallant1988}, as $\Theta$ is compact and the log likelihood is continuous in $\theta$ and measurable in $x_{i}$. The consistency of the ML estimator, $\hat{\theta}_{n}(\hat{f}_1) \stackrel{as}{\to} \theta_{0}$, is obtained by \citet[Theorem 3.4]{White1994} or \citet[Theorem 3.3]{Gallant1988}. Below, we note that we satisfy the sufficient conditions of uniform convergence of the log likelihood function
\begin{equation*}
\sup_{\theta \in \Theta} |\hat L_{n}(\theta) - L_{\infty}(\theta)| \stackrel{as}{\to} 0 \ \forall \ \hat{f}_1\in \mathcal{F} \ \text{ as } \ n\to \infty,
\end{equation*} 
and the identifiable uniqueness of the maximizer $\theta_{0} \in \Theta$ introduced in \cite{White1994},
\begin{equation*}
\sup_{\theta: \| \theta - \theta_{0} \| > \epsilon} L_{\infty}(\theta) < L_{\infty}(\theta_{0}) \ \forall \ \epsilon >0.
\end{equation*} 
The uniform convergence of the criterion is obtained since, by norm sub-additivity, we can split the log likelihood as follows
\begin{equation}
\label{eq2pt2wq}
\sup_{\theta \in \Theta} |\hat L_{n}(\theta) - L_{\infty}(\theta)| \leq \sup_{\theta \in \Theta} |\hat L_{n}(\theta) - L_{n}(\theta)| + \sup_{\theta \in \Theta} |  L_{n}(\theta) - L_{\infty}(\theta)|.
\end{equation}
The first term on the right-hand-side of (\ref{eq2pt2wq}) vanishes if $  |\hat l_{i}(\theta) - l_{i}(\theta)| \stackrel{as}{\to} 0$ since
\begin{equation*}
|\hat L_{n}(\theta) - L_{n}(\theta)| \leq \frac{1}{n}\sum^{n} |\hat l_{i}(\theta) - l_{i}(\theta)| \stackrel{as}{\to} 0,
\end{equation*}
and we have that 
\begin{equation*}
\sup_{\theta \in \Theta} |\hat l_{i}(\theta) - l_{i}(\theta)| \leq \sup_{\theta \in \Theta}\sup_{f} |\nabla(x_{i},f,\theta)|  \cdot \sup_{\theta \in \Theta}|\hat f_{i}(\theta) - f_{i}(\theta)| \stackrel{as}{\to} 0 \quad \forall \ \hat{f}_1\in \mathcal{F} \quad \text{ as } \quad n\to \infty,
\end{equation*}
where $\sup_{\theta \in \Theta}|\hat f_{i}(\theta) - f_{i}(\theta)|\stackrel{as}{\to} 0 $ follows from the invertibility of the filter (proved in Proposition \ref{prop:invertibility}) and the product vanishes by the bounded logarithmic moment of the score $\mathrm{E} [ \ln^{+}\sup_{f}|\nabla(x_{i},f)| ]<\infty$ (see Lemma 2.1 in \citealt{Straumann2006}). 
The logarithmic moment $\mathrm{E}[\ln^{+}\sup_{f}|\nabla(x_{i},f)|]<\infty$ follows as
\begin{equation*}
\begin{aligned}
\mathrm{E} \left[ \ln^{+} |s (0, \hat{f}_i)| \right] &=  \mathrm{E} \left[ \ln^{+} \left|
\frac{\exp(\hat{f}_{i})(\pi - 1) }{(\alpha \exp(\hat{f}_{i}) + 1) \left( 1 + \pi (\alpha \exp(\hat{f}_{i}) + 1)^{\alpha^{-1}} - \pi \right)} \right| \right] < \infty, \\
\mathrm{E}_{x_{i}>0} \left[ \ln^{+}| s (x_{i}, \hat{f}_i)| \right] &= \left| \frac{x_i - \exp(\hat{f}_{i})}{\alpha \exp(\hat{f}_{i}) + 1} \right| < \infty \quad \text{for }  x_i >0.
\end{aligned}
\end{equation*}
Note that since we use unit scaling in Lemma \ref{theo:consistency}, we have that $\nabla(x_{i},f)=s\nabla(x_{i},f)$.
The uniform convergence of the second term on the right-hand-side of (\ref{eq2pt2wq})
\begin{equation*}
\sup_{\theta \in \Theta} |  L_{n}(\theta) - L_{\infty}(\theta)| \stackrel{as}{\to} 0 \quad \forall \  \hat{f}_1\in \mathcal{F} \quad \text{ as } \quad n\to \infty,
\end{equation*}
follows by application of the ergodic theorem for separable Banach spaces in \citet{Rao1962}. We note that the $\{L_{n}(\cdot)\}_{t \in \mathbb{N}}$ has strictly stationary and ergodic elements as it depends on the limit strictly stationary and ergodic filter  taking values in the Banach space of continuous functions $\mathbb{C}(\Theta,\mathbb{R})$ equipped with sup norm. We also note that $L_{n}(\cdot)$ has one bounded moment since $\mathrm{E}[ L_{n}(\theta)] \leq \frac{1}{n}\sum^{n} \mathrm{E}[l_{i}(\theta)]<\infty$. 
In particular, the bounded moment for the log likelihood holds trivially if the data has a bounded moment $\mathrm{E}[x_{i}]<\infty$ since  $\ln \ell_{i}(x_{i},\theta)$ is bounded in $\mu_i$ and bounded by a linear function in $x_{i}$,
\begin{equation*}
\begin{aligned}
\ell_{i}(0,\theta) &= \ln \emph{P}[X_i = 0 | \hat{f}_{i}(\theta),\theta] \\
&= \ln \left( \pi + (1 - \pi) \left( \frac{\alpha^{-1}}{\alpha^{-1} + \exp (\hat f_{i}(\theta))} \right)^{\alpha^{-1}} \right), \\
\ell_{i}(x_{i},\theta) &= \ln \emph{P}[X_i = x_i | \hat{f}_{i}(\theta),\theta] \\
&= \ln (1 - \pi) + \ln  \frac{\Gamma (x_i + \alpha^{-1})}{\Gamma (x_i + 1) \Gamma (\alpha^{-1})} \\
& \quad + \frac{1}{\alpha} \ln  \left( \frac{\alpha^{-1}}{\alpha^{-1} + \exp (\hat f_{i}(\theta))} \right) + x_{i} \ln  \left( \frac{\exp (\hat f_{i}(\theta))}{\alpha^{-1} + \exp (\hat f_{i}(\theta))} \right) \quad \text{for } x_i>0.
\end{aligned}
\end{equation*}

The identifiable uniqueness (see e.g.\ \citealp{White1994}) 
follows from the compactness of $\Theta$,  the assumed  uniqueness of $\theta_{0}$, and the continuity of the limit likelihood function $\mathrm{E}[\ell_{i}(\theta)]$ in $\theta \in \Theta$.
\medskip

\noindent \textit{Proof of Lemma \ref{theo:normality}:} 

This proof follows \citet[Theorem 4.16]{Blasques2022}.	In particular, we obtain the asymptotic normality using the usual expansion argument found e.g.\ in \citet[Theorem 6.2]{White1994} by establishing: 
\begin{itemize}
\item[(i)] The consistency of $\hat{\theta}_{n} \stackrel{as}{\to} \theta_{0} \in \mathrm{int}(\Theta)$, which follows immediately by Lemma \ref{theo:consistency}.
\item[(ii)] The as~twice continuous differentiability of $L_{n}(\theta,\hat{f}_1)$ in $\theta \in \Theta$, which holds trivially for our zero-inflated score model.
\item[(iii)] The asymptotic normality of the score, which can be shown to hold by verifying that,
\begin{equation}
\label{eqsclt}
\sqrt{n}\frac{\partial L_{n}(\theta_{0})}{\partial \theta}   \stackrel{d}{\to} N(0,\mathcal{I}(\theta_{0})\big) \ \text{ as } \ n \to \infty, \ 
\end{equation}
and
\begin{equation}
\label{eqins}
\sqrt{n}	\Big| \frac{\partial\hat{L}(\theta_{0})}{\partial \theta} -  \frac{\partial L (\theta_{0})}{\partial \theta} \Big| \stackrel{as}{\to} 0 \ \text{ as } \ n \to \infty.
\end{equation}
The asymptotic normality in (\ref{eqsclt}) is obtained by  application of a central limit theorem for martingale difference sequences to the score, after noting that the score 
\begin{equation*}
\frac{\partial L_{n}(\theta_{0})}{\partial \theta}= \frac{1}{n} \sum^{n} 
\left(\frac{\partial   \ell_{i}(x_{i},\theta_{0})}{\partial \theta} + \frac{\partial    \ell_{i}(x_{i},\theta_{0})}{\partial f_{i}} \frac{\partial   f_{i}(\theta_{0})}{\partial \theta} \right).	 
\end{equation*}
has two bounded moments. In particular, 
\begin{equation*}
\mathrm{E}	\left[ \left\| \frac{\partial L_{n}(\theta_{0})}{\partial \theta} \right\|^{2} \right] \leq \mathrm{E} \left[ \left\| \frac{\partial    \ell_{i}(x_{i},\theta_{0})}{\partial \theta} \right\|^{2}  \right]+ \mathrm{E} \left[ \left\| \frac{\partial \ell_{i}(x_{i},\theta_{0})}{\partial f_{i}} \frac{\partial   f_{i}(\theta_{0})}{\partial \theta} \right\|^{2} \right] < \infty,	 
\end{equation*}
where the bounds
\begin{equation*}
\mathrm{E} \left[ \left\| \frac{\partial  \ell_{i}(x_{i},\theta_{0})}{\partial \theta} \right\|^{2} \right] < \infty \quad \text{and} \quad 
\mathrm{E} \left[ \left\| \frac{\partial    \ell_{i}(x_{i},\theta_{0})}{\partial f_{i}} \frac{\partial   f_{i}(\theta_{0})}{\partial \theta} \right\|^{2} \right] < \infty,
\end{equation*}
hold, for example, under the assumption that 
\begin{equation*}
\mathrm{E} \left[ \left\| \frac{\partial  \ell_{i}(x_{i},\theta_{0})}{\partial f_{i}} \right\|^{4} \right] <\infty \quad \text{and} \quad \mathrm{E} \left[ \left\| \frac{\partial  \ell_{i}(x_{i},\theta_{0})}{\partial \theta} \right\|^{4} \right] < \infty;
\end{equation*}
by a generalized Holder's inequality as used e.g.~in \citet{Blasques2022}.
For the negative binomial model it is easy to see for example that the four bounded moments for score term $\partial \ell_{i}(x_{i},\theta_{0})/\partial f_{i}$ can be obtained if the data has four bounded moments, $\mathrm{E}|x_{i}|^{4}<\infty$, by noting that 
\begin{equation*}
\begin{aligned}
\mathrm{E} \left[ \sup_{\theta \in \Theta} \| s(0,\hat{f}_{i},\theta)\|^{4} \right] &\leq \sup_{\mu}  \sup_{\theta \in \Theta} \| s(0,\hat{f}_{i},\theta)\|^{4} \\
&=  \sup_{\mu}  \sup_{\theta \in \Theta} \left| 	(\pi - 1)   \exp(\hat{f}_i)  (\alpha \exp(\hat{f}_{i}) + 1)^{-1} \left( 1 + \pi (\alpha \exp(\hat{f}_i) + 1)^{\alpha^{-1}} - \pi \right)^{-1} \right|^{4} \\
&< \infty, \\
\end{aligned}
\end{equation*}
since $s(0,\hat{f}_{i},\theta)$ is uniformly bounded in $\hat{f}_{i}$. Furthermore, by application of the so-called $c_{n}$-inequality, there exists a finite constant $k$ such that,
\begin{equation*}
\begin{aligned}
\mathrm{E}_{x_{i}>0} \left[ \sup_{\theta \in \Theta}| s(x_{i},\hat{f}_{i},\theta)|^{4} \right] &= \mathrm{E}_{x_{i}>0} \left[ \sup_{\theta \in \Theta} \left|   x_i - \exp(\hat{f}_i)(\alpha \exp(\hat{f}_{i}) + 1)^{-1} \right|^{4} \right] \\
&\leq k \sup_{\theta \in \Theta}\frac{1}{\alpha} \mathrm{E}_{x_{i}>0} [ x_i^{4} ] + k|\alpha^{-1}|^{4} \\
&< \infty. \\
\end{aligned}
\end{equation*}

Additionally, following the argument of \citet[Theorem 4.14]{Blasques2022}  and \citet[Lemma 2.1]{Straumann2006},  the as~convergence in (\ref{eqins}) follows by the invertibility of the filter and its derivatives. The invertibility of the first derivative process  can be verified by applying Theorem 2.10 in \cite{Straumann2006}. This theorem is analogue to Theorem 3.1 of \cite{Bougerol1993}, also used in the proof of Proposition \ref{prop:invertibility} above, but it applies to perturbed stochastic sequences. For example, the updating equation for derivative process $\partial f_{i}/\partial c=\partial \hat{f}_{i}/\partial c$ takes the form
\begin{equation*}
\begin{aligned}
\frac{\partial \hat{f}_{i+1}}{\partial c } &= 1 + b   \frac{\partial \hat{f}_{i}}{\partial c} 	+ \frac{\partial s(x_{i},\hat{f}_{i})}{\partial \hat{f}_{i}} \frac{\partial \hat{f}_{i}}{\partial c} = 1 + \left(b + \frac{\partial s(x_{i},\hat{f}_{i})}{\partial \hat{f}_{i}} \right) \frac{\partial \hat{f}_{i}}{\partial c} .
\end{aligned}
\end{equation*}
Hence, by application of Theorem 2.10 in \cite{Straumann2006}, the invertibility of this filter is ensured by (a) the invertibility of the filter $\{\hat{f}_{i}\}_{i\in \mathbb{N}}$ (shown in Proposition \ref{prop:invertibility}); (b) the  contraction condition $\mathrm{E} [ \ln |b  	+  \partial s(x_{i},\hat{f}_{i})/\partial \hat{f}_{i}| ]<0$; and a logarithmic moment for $\partial^{2} s(x_{i},\hat{f}_{i})/\partial \hat{f}_{i}^{2}$.

\item[(iv)] The uniform convergence of the Hessian, is obtained through the invertibility of the filter and its derivative processes. In particular, a sufficient condition is for 
the first and second derivatives of the filtering process to converge almost surely, exponentially fast, to a limit stationary and ergodic sequence, 
\begin{equation*}
\left\| \frac{\partial \hat f_{i}(\theta_{0})}{\partial \theta} - \frac{\partial f_{i}(\theta_{0})}{\partial \theta} \right\| \stackrel{eas}{\to} 0 \quad \text{and} \quad \sup_{\theta \in \Theta} \left\| \frac{\partial^{2} \hat f_{i}(\theta)}{\partial \theta \partial \theta'} - \frac{\partial^{2} f_{i}(\theta)}{\partial \theta \partial \theta'} \right\| \stackrel{eas}{\to} 0 \quad \text{as}\quad  i \to \infty,
\end{equation*}
with four bounded moments
\begin{equation*}
\mathrm{E} \left[ \left\| \frac{\partial f_{i}(\theta_{0})}{\partial \theta} \right\|^{4} \right] <\infty
\quad \text{and} \quad \mathrm{E} \left[ \sup_{\theta \in \Theta} \left\| \frac{\partial^{2} f_{i}(\theta)}{\partial \theta \partial \theta'} \right\|^{4} \right] < \infty.
\end{equation*}
and to have logarithmic moments for cross derivatives, 
\begin{equation*}
\mathrm{E} \left[ \sup_{\theta \in \Theta} \left\| \frac{\partial^{2} \ell_{i}(x_{i},\theta)}{\partial f_{i} \partial \theta'} \right\| \right] <\infty, \quad \mathrm{E} \left[ \sup_{\theta \in \Theta} \left\| \frac{\partial^{2}  \ell_{i}(x_{i},\theta)}{\partial f_{i}^{2} } \right\| \right] <\infty \quad \text{and} \quad \mathrm{E} \left[ \sup_{\theta \in \Theta} \left\| \frac{\partial^{2}  \ell_{i}(x_{i},\theta)}{\partial \theta \partial \theta'} \right\| \right] <\infty;
\end{equation*}
and also for the third-order derivatives of the log likelihood to have a uniform logarithmic bounded moment,
\begin{equation*}
\mathrm{E} \left[ \ln^{+}\sup_{\theta \in \Theta} \left\| \frac{\partial^{3}  \ell_{i}(x_{i},\theta_{0})}{\partial f_{i}^{2} \partial \theta'} \right\| \right] <\infty, \quad \mathrm{E} \left[ \ln^{+}\sup_{\theta \in \Theta} \left\| \frac{\partial^{3}  \ell_{i}(x_{i},\theta_{0})}{\partial f_{i}^{3} } \right\| \right] <\infty.
\end{equation*}
\begin{equation*}
\text{and} \quad \mathrm{E} \left[ \ln^{+}\sup_{\theta \in \Theta} \left\| \frac{\partial^{3}  \ell_{i}(x_{i},\theta_{0})}{\partial \theta \partial \theta' \partial f} \right\| \right] <\infty;
\end{equation*}
Then by application of the  ergodic theorem for separable Banach spaces in \citet{Rao1962} to the limit Hessian (see also \citealt{Blasques2022} and \citealt[Theorem 2.7]{Straumann2006} for additional details), we have,
\begin{equation}
\label{equcinf}
\sup_{\theta \in \Theta} \left\| \frac{\partial^{2} L_{n}(\theta)}{\partial \theta \partial \theta'} - \mathrm{E} \left[ \frac{\partial^{2} \ell_{i}(\theta)}{\partial \theta \partial \theta'} \right] \right\|  \stackrel{as}{\to} 0 \quad \text{as } n \to \infty.
\end{equation}
\item[(v)] The non-singularity of the limit $L_{\infty}''(\theta) = \mathrm{E}[\ell_{i}''(\theta)] = \mathcal{I}(\theta)$ follows by the uniqueness of $\theta_{0}$ and the independence of derivative processes (\citealt[Theorem 2.7]{Straumann2006}).
\end{itemize}

\section{Model Evaluation}
\label{app:eval}

It is well know that ranking models based on their expected log-likelihood $\mathrm{E}[\ell_{i}(\theta_{0})]$ evaluated at the best (pseudo-true) parameter $\theta_{0}$ is equivalent to  model selection based on minimizing the expected Kullback-Leibler divergence between the true distribution of the data and the model-implied distribution.  The sample log-likelihood is however an asymptotically biased estimator of the expected log likelihood. Under restrictive conditions, \cite{Akaike1973,Akaike1974} showed that the bias is approximately given by the number of parameters of the model $\dim(\theta)$. Since then, the AIC has been shown to consistently rank models according to the Kullback-Leibler divergence under considerably weaker conditions (\citealt{Sin1996,Konishi2008}). Unfortunately, model specification and identification issues still exert a  strong influence over the performance of in-sample information criteria.

For this reason, it could be interesting to consider criteria based on a validation sample. Lemma \ref{cmrv} highlights that log-likelihood based on an independent validation sample of $m$ observations, $n \hat{L}_{m}(\hat{\theta}_{n})$, is asymptotically  unbiased for $n \mathrm{E}[\ell_{i}(\theta_{0})]$. A proof can be found in \cite{Andree2017}\footnote{For time-series data with  fading memory, a burn-in period between the estimation and the validation samples can be the approximate independence between the two samples. Proofs then rely on expanding estimation, burn-in and validation samples.}.

\begin{lemma}
\label{cmrv}
Let the conditions of Lemma \ref{theo:consistency} hold. Then $\lim_{n,m\to \infty}  \mathrm{E} \left[ n \hat{L}_{m}(\hat{\theta}_{n})-n\mathrm{E}[\ell_{i}(\theta_{0})] \right] =  0$.
\end{lemma}

Lemma  \ref{propdbts} uses a Diebold-Mariano test statistic \citep{Diebold1995} to test for differences in log-likelihoods across different models obtained from the validation sample (see \citealp{Andree2017}, for a proof). This test is also known as a logarithmic scoring rule, see e.g.\ \cite{Diks2011,Amisano2007,Bao2007b}. Given two models, A and B, let $\tilde{\ell}_{i}^{\text{A}}(\theta_{0}^{\text{A}})$ and $ \tilde{\ell}_{i}^{\text{B}}(\theta_{0}^{\text{B}})$ denote their respective log-likelihood contributions at a certain time $i$ (in the validation sample) evaluated at each model's pseudo-true parameter. Define the log-likelihood difference 
\begin{equation*}
D_i^{A,B}:=\tilde{\ell}_{i}^{\text{A}}(\theta_{0}^{A}) -  \tilde{\ell}_{i}^{\text{B}}(\theta_{0}^{B}).
\end{equation*}
Finally, define the Diebold-Mariano test statistic 
\begin{equation*}
\text{\emph{DM}}_{m,n} = \sqrt{m} \frac{\mu_{D}^{A,B}}{\sigma_{D}^{A,B}}, \qquad \mu_{D}^{A,B} = \frac{1}{m} \sum_{i = n+1}^{n+m} D_i^{A,B}, \qquad \sigma_{D}^{A,B} = \sqrt{ \frac{1}{m-1} \sum_{i = n+1}^{n+m} \left( D_i^{A,B} - \mu_{D}^{A,B} \right)^2 }.
\end{equation*}

\begin{lemma}[Validation-Sample Test]
\label{propdbts}
Let Lemma \ref{theo:consistency} hold for both models A and B, such that  $\hat{\theta}_{n}^{A} \stackrel{as}{\to}\theta_{0}^{A}$ and $\hat{\theta}_{n}^{B} \stackrel{as}{\to}\theta_{0}^{B}$ as $n \to \infty$. Then  we have that 
\begin{equation*}
\text{\emph{DM}}_{m,n} \stackrel{d}{\to}\mathcal{N}(0,1) \quad \text{as} \ \ n,m \to \infty,
\end{equation*}	
under the null hypothesis \emph{H}$_{0}: \mathrm{E} [D_m^{A,B}]=0$, 
where  $\sigma_{D}^{A,B}$ is a consistent estimator of the standard deviation of  $D_m^{A,B}$.
If $\mathrm{E}[D_m^{A,B}]>0$ then $\text{\emph{DM}}_{m,n} \to  \infty$ as $n,m \to \infty$. Finally, if $\mathrm{E}[D_m^{A,B}]<0$, then $\text{\emph{DM}}_{m,n} \to - \infty$.
\end{lemma}

\section{Generalized Gamma Distribution}
\label{app:gengamma}

The generalized gamma distribution is a continuous probability distribution and a three-parameter generalization of the two-parameter gamma distribution \citep{Stacy1962}. It also contains the exponential distribution and the Weibull distribution as special cases. It uses the scale parameter $\beta$ and two shape parameters $\theta$ and $\varphi$. The probability density function is
\begin{equation*}
\label{eq:gengammaDensity}
p(x | \beta, \theta, \varphi) = \frac{1}{\Gamma \left( \theta \right) } \frac{\varphi}{\beta} \left( \frac{x}{\beta} \right)^{\theta \varphi - 1} e^{- \left( \frac{x}{\beta} \right)^{\varphi}} \quad \text{for } x \in (0, \infty).
\end{equation*}
The expected value and variance is
\begin{equation*}
\label{eq:gengammaMoments}
\begin{aligned}
\mathrm{E}[X] &= \beta \frac{\Gamma \left(\theta + \varphi^{-1} \right)}{\Gamma \left( \theta \right)}, \\
\mathrm{var}[X] &= \beta^2 \frac{\Gamma \left( \theta + 2 \varphi^{-1} \right)}{\Gamma \left( \theta \right)} - \left( \beta \frac{\Gamma \left(\theta + \varphi^{-1} \right)}{\Gamma \left( \theta \right)} \right)^2. \\
\end{aligned}
\end{equation*}
The score vector is
\begin{equation*}
\label{eq:gengammaScore}
\nabla (x; \beta, \theta, \varphi) = \begin{pmatrix}
\varphi \beta^{-1} \left( x^{\varphi} \beta^{-\varphi} - \theta \right) \\
\varphi \ln \left(x \beta^{-1} \right) - \psi_0( \theta ) \\
\theta \ln \left( x \beta^{-1} \right) -  x^{\varphi} \beta^{-\varphi}  \ln \left( x \beta^{-1} \right) + \varphi^{-1} \\
\end{pmatrix} 
\quad \text{for } x \in (0, \infty).
\end{equation*}
Special cases of the generalized gamma distribution include the gamma distribution for $\varphi = 1$, the Weibull distribution for $\theta = 1$ and the exponential distribution for $\theta = 1$ and $\varphi = 1$.

\end{document}